\documentclass[12pt]{article}

\usepackage{amsmath,geometry,amsthm,amssymb,natbib,verbatim,upref,paralist,indentfirst,verbatim,graphicx,pdfpages,caption,hyperref}

\usepackage{subcaption}
\usepackage{graphbox,graphicx}
\usepackage{booktabs}
\usepackage{multirow}
\usepackage[section]{placeins}
\newtheorem{proposition}{Proposition}

\newtheorem{corollary}{Corollary}

\newtheorem{remark}{Remark}

\usepackage[T1]{fontenc}
\usepackage[utf8]{inputenc}
\usepackage{authblk}
\usepackage{tgheros}
\begin{document}

\title{Strategic Bidding in Knapsack Auctions\thanks{This research is funded by The Ethereum Foundation Academic Grant Program no. FY22-0682. We would like to thank various seminar participants for insightful comments. All errors remain our own.  }}

\author{Peyman Khezr\thanks{Royal Melbourne Institute of Technology. Email: peyman.khezr@rmit.edu.au.  }
\space\space Vijay Mohan \thanks{Royal Melbourne Institute of Technology. Email: vijay.mohan@outlook.com.} 
\space\space Lionel Page \thanks{School of Economics, University of Queensland. Email: lionel.page@uq.edu.au.}}


\date{}
\maketitle

\begin{abstract}

This paper examines knapsack auctions as a method to solve the knapsack problem with incomplete information, where object values are private and sizes are public. We analyze three auction types—uniform price (UP), discriminatory price (DP), and generalized second price (GSP)—to determine efficient resource allocation in these settings. Using a Greedy algorithm for allocating objects, we analyze bidding behavior, revenue and efficiency of these three auctions using theory, lab experiments, and AI-enriched simulations. Our results suggest that the uniform-price auction has the highest level of truthful bidding and efficiency while the discriminatory price and the generalized second-price auctions are superior in terms of revenue generation. This study not only deepens the understanding of auction-based approaches to NP-hard problems but also provides practical insights for market design.

\end{abstract}

\vspace{1ex}
\noindent\textit{Keywords}: Knapsack problem; auctions; experiment; Q-learning.\\[1ex]
\noindent \textit{JEL Classification}: D44, D82, C91, C63.


\section{Introduction}





Auctions play a pivotal role in allocating goods and services in various markets. In particular, multi-unit auctions are instrumental in discovering prices and allocating resources in environments where demands are diverse and quantities of goods vary. In most multi-unit auctions, buyers demand bundles of goods, which are typically divisible. However, there are many instances where the demand is inflexible, and the buyer would only derive positive utility from acquiring the entire bundle, receiving zero utility otherwise. In such situations, the principal is essentially dealing with the problem of allocating a fixed supply of goods (or space) to a group of agents with varying and non-adjustable demands (sizes). A decision problem following this general structure is referred to as the \textit{knapsack problem} \citep{kellerer2004multidimensional}.\footnote{The intuition behind this decision problem is often couched in the context of an individual, say a mountaineer, seeking to fill a knapsack with a set of useful objects. The mountaineer places some value on carrying each object, and would ideally like to carry the entire set. However the mountaineer faces a capacity constraint, such as the total volume of objects that can be fit in the knapsack, or the total weight that the mountaineer can carry; the allowable volume or weight represents the scarce resource. \cite{kellerer2004multidimensional} present an in-depth discussion of the knapsack problem variations; see \cite{bartholdi2008knapsack} for an intuitive introduction.} In this paper, our focus is on \textit{knapsack auctions}: scenarios where the decision about which objects to fit into the knapsack is determined through an auction mechanism. Knapsack auctions have a variety of potential applications in real-world markets, including advertising spot allocation, spectrum allocation, cloud computing resource allocation, electricity market grid capacity allocation, event ticketing and seating, and container or freight space allocation, to name a few.\footnote{More recently, the knapsack problem has found considerable use in blockchains, where transactions of different sizes generating fees of different amounts are selected for inclusion into a block of fixed capacity \citep{mohan2024blockchains}.}

In this paper, we investigate the use of practical knapsack auctions to tackle the knapsack problem with incomplete information. Our analysis features a `seller' of scarce space in the knapsack and `buyers' aiming to secure a spot within it, effectively translating the knapsack problem into a market design problem. We assume that the values of the objects are known exclusively to their owners, while the sizes are public information.\footnote{Indeed, it is intuitive to think that decisions about filling the knapsack cannot be made if the sizes of objects cannot be observed, as the feasible alternatives cannot be known with certainty. Consequently, in many realistic and tractable situations, private information exists only for a single parameter: the values.} In particular, we look at three multi-unit auctions for selling knapsack space to buyers: first, the uniform price (UP) auction, where buyers pay amounts based on the per-unit bid of the highest losing bidder; second, the discriminatory price (DP) auction, where buyers pay their own bid for getting their object included in the knapsack; and third, the generalized second price (GSP) auction, where a buyer pays an amount based on the per-unit bid of the next highest bidder.\footnote{The primary reason for selecting these three auction types is their widespread application in real-world markets. Uniform-price and discriminatory auctions are extensively utilized by governments to allocate Treasury bills and emission permits. Similarly, the generalized second-price auction is employed by search engines for advertisement spot allocation.} Our analysis explores this from three different perspectives: first, theoretical; second, experimental, with human participants as buyers in laboratory experiments; and third, simulations using buyers with artificial intelligence (AI).

Our key results show that the uniform price auction is the unique dominant strategy incentive compatible mechanism in knapsack auctions. However, it generates significantly lower revenue than the discriminatory and generalized second-price auctions. The latter two yield similar revenues, with the generalized second-price auction performing best overall in terms of both revenue and efficiency. These results are borne out in both the lab experiments with human bidders and in our AI simulations. The simulations allow us to examine a much larger strategy space and number of repetitions than is feasible in the lab.

This paper contributes to several strands of literature, including auctions, market mechanisms, and practical solutions for addressing NP-hard problems. Our experimental methodology is augmented by AI-enriched simulations to explore economic behavior evolution in complex settings characterized by large strategy sets and several agents. Utilizing AI simulations enables us to circumvent experimental constraints, facilitating extended learning across more episodes than is feasible with human participants. The applicability of this innovative method extends beyond this paper and can be adopted in other contexts with similar attributes.

The knapsack problem has a direct solution when objects are divisible: fill the knapsack prioritizing objects by their value per unit size until full. This method known as the \textit{Greedy algorithm} \citep{dantzig1957discrete}, results in the maximum value sum. However, with indivisible objects, the complexity increases significantly, rendering it an NP-hard problem without a polynomial-time solution \citep{kellerer2004multidimensional}. To see how the Greedy algorithm can fail, consider a knapsack with a 10-pound capacity and two items: one weighing 1 pound worth \$1, and another weighing 9.9 pounds worth \$9. Using the greedy algorithm, which selects items based on value per pound, the 1-pound item is chosen, leaving the knapsack nearly empty with \$1 in value. In contrast, choosing the second item would nearly fill the knapsack, showcasing \$9 in value, despite its lower value per pound, illustrating the algorithm's limitations in this scenario. Nevertheless, the Greedy algorithm serves as a useful heuristic, and the inefficiency arising from employing the Greedy algorithm has known and well-understood bounds that we elaborate upon presently. As such, in this paper, we examine the interesting case of indivisible objects and employ the Greedy algorithm as the allocation rule to simplify the computational problem.

While there is some literature on knapsack auctions, by and large the focus has been on a broad mechanism design approach that seeks a truthful equilibrium \citep{aggarwal2006knapsack}. In computationally complex problems such as knapsack auctions or combinatorial auctions, when the optimal outcome is replaced with a more tractable approximation algorithm, the VCG mechanism is no longer necessarily truthful \citep{lehmann2002truth, mu2008truthful, nisan2007computationally, akbarpour2022investment}. As the social optimum is computationally difficult to solve, so too is the corresponding VCG payments; in other words, calculating the VCG payments for the socially optimum allocation is computationally hard. Given that the optimal selection of bids in a knapsack auction is NP-hard, in this paper we use an approximation algorithm – the Greedy algorithm – which yields a sub-optimal outcome. So, one of the first questions that arises is: given a Greedy allocation, is there an auction that implements a truthful equilibrium? The answer is yes; in single parameter domains, if the allocation rule is monotone, there exists a unique pricing rule that is dominant strategy incentive compatible \citep{myerson1981optimal, nisan2007introduction,roughgarden2016twenty}. In our paper, each agent $i$ receives a private value, $v_i$, from winning the auction and securing space in the knapsack, and 0 otherwise. In other words, the single parameter $v_i$ describes the agent’s value in all winning alternatives. Moreover, it is readily verified that the Greedy algorithm is monotone. Consequently, every winning bidder $i$ pays a price that equals a critical value below which $i$ loses and above which $i$ wins \citep{nisan2007introduction, roughgarden2016twenty}. 


Knapsack auctions have also been examined in the context of advertising on internet search engines \citep{aggarwal2006knapsack}. As it happens, internet search engines like Google utilize neither the UP nor the DP auction; rather, advertisements in internet searches are sold through generalised second price (GSP) auctions, which do no not have a truthful equilibrium \citep{Edelman2007}. We show that the absence of a truth-telling equilibrium for GSP holds true in the context of a knapsack auction as well. 


There are various studies that have run experiments to see how human subjects can solve complex problems like the knapsack problem \citep{bossaerts2020computational, murawski2016humans}. For example, \cite{murawski2016humans} suggest that there is a counter-intuitive trend in problem-solving efforts: although participants generally invested more effort into tasks that demanded higher computational resources, their efficacy in solving these tasks was inversely affected, showing a decrease in success rates. While to our knowledge there is no study that investigates a knapsack auction experimentally, there are several that examine, in different contexts, the payment rules we focus on in this paper. For instance, \cite{bae2019experimental} study GSP auctions in the context of advertising positions. They examine two distinct click-through rates (CTRs) within both static complete and dynamic incomplete information frameworks. In contrast to the equilibrium suggested by the VCG mechanism, they find that subjects' bids were consistently higher. \cite{sade2006competition} is another example of an experimental study that compares the DP and the UP auctions, using treasury auctions as the setting. They find that DP auctions are more vulnerable to collusion compared the UP auction, unlike what theory predicts.\footnote{For a comprehensive literature review of multi-unit auctions with homogeneous goods see \cite{Khezr2022}. For spectrum auctions see \cite{bichler2017handbook}.}

More recently there have been a few papers that have investigated the use of artificial intelligence in auction design. For example, \cite{banchio2022artificial} utilizes a Q-learning algorithm to simulate the first-price and second-price single unit auctions. They show in a repeated auction environment that the first-price auction may result in tacit-collusive outcomes, unlike the second-price auction. \cite{calvano2020artificial} is another example where authors used Q-learning algorithms to study price competition between oligopolies in a repeated game. As outlined in Section \ref{AI}, we employ a more advanced Q-learning algorithm to address the complexities that arise due to the larger strategy set available to each agent and the possibility of multiple winners in each auction, each with different rewards.

\section{Theoretical Model}\label{theo}

An owner (the seller) of a knapsack space with a capacity $K>0$ would like to sell the space of its knapsack to a set of $N=\{1,2,...,n\} $ potential buyers. Each buyer $i$ owns an object that would occupy $k_i<K$ space in the knapsack. We assume all objects' sizes are public information and to avoid trivial cases, we presume that $\sum\limits_{i\in N} k_i >K$. 

Each buyer $i$ would realize a value $v_i$ if their object is packed in the knapsack and would otherwise receive zero. Suppose $v_i$ is privately known by buyer $i$, however, it is commonly known that values are distributed according to some distribution function $F(.)$ which is continuous and twice differentiable with $f<\infty$.

Bidding in a knapsack auction starts with each bidder $i$ submitting a bid $B_i$ for their object with size $k_i$. The auctioneer uses the Greedy algorithm to pack the knapsack, which involves ranking objects based on the per unit bid from the highest to the lowest. We denote per unit bid of bidder $i$ as $b_i$, and without loss of generality assume that $b_1>b_2>...>b_n$. The auctioneer fills the knapsack starting from the highest per unit bid until there is no more space for the next object in the line.\footnote{There are two technical points worth noting. First, this assumes that the auctioneer stops as soon as the \textit{next} per unit bid corresponds to an object that cannot be packed into the knapsack. There may be subsequent (even lower) per unit bids that involve smaller objects that could have been packed into the knapsack. This leads to some ambiguity as to what the highest losing bid is for a uniform price auction, so we assume that the auctioneer prefers not to deal with added complexity that arises from this ambiguity. Second, the Introduction provided an example of a perverse case where the Greedy algorithm fails. To rule out such cases, the auctioneer would need to pick either the solution from the Greedy algorithm or the highest bidder, whichever yields a better outcome. In this paper, the experiments and AI simulations reported in Section 4 and 5 are based on values and sizes where this problem does not occur. Consequently, we focus on the outcome of the Greedy algorithm.} The next step in the auction is the payment transfer for those objects that are packed. In this paper, we consider three possible payment rules: the uniform-price (UP), the discriminatory price (DP) and the generalized second-price (GSP). 

In the DP auction, every bidder $i$ whose object is packed pays $B_i$ to the auctioneer. In the GSP auction, each bidder $i$ whose object is packed (following a bid $B_i$) will pay an amount that is based on the per unit bid submitted by the \textit{next} bidder, that is, $i$ will pay $k_i b_{i+1}$. Finally, in the UP auction every bidder whose object is packed pays the per unit bid submitted by the first person in line whose object has not been packed. Formally if we denote the highest per unit bid that has not been packed by $b_j$, every bidder $i$ with packed objects would pay $k_i b_j$.

 \subsection{The Uniform-Price (UP) auction}
 
We begin with the uniform-price auction as the benchmark case in this study. The uniform-price auction is known for untruthful bids and demand reduction in the literature of multi-unit auctions \citep{Krishna2009}. However, for our setting, we demonstrate that it is the unique dominant-strategy incentive-compatible (DSIC) payment rule of the knapsack auction game. In the uniform-price format all packed bidders pay the same per unit price equal to the highest per unit bid which is not packed. The expected payoff of a bidder $i$ becomes,

\begin{equation}
\pi_i= (v_i - k_ib_{j}) Pro(k_i \: \text{being packed})
\end{equation}

\begin{proposition}\label{proup}
The UP auction has the unique dominant-strategy incentive-compatible equilibrium of the knapsack auction game.
\end{proposition}
 
\begin{proof}
We first show the incentive compatible property of the UP auction. Denote $\mathbf{b}^*$ as the vector of equilibrium bids that is incentive compatible. We first show this equilibrium exists in dominant strategies. If bidder $i$ follows any strategy $b'_i > \frac{v_i}{k_i}$ there are three possibilities. First their bid is among the winners and their per unit value is higher than the highest losing bid. In this case they would receive the same payoff as the case where the bid is $b_i = \frac{v_i}{k_i}$. Second their bid is among the winners and their per unit value is less than the highest losing bid. In this case the would receive a negative payoff while bidding $b_i = \frac{v_i}{k_i}$ results in zero payoff. Finally if their bid is among the losers they would receive zero similar to a bid equal to $b_i = \frac{v_i}{k_i}$. Therefore $b_i = \frac{v_i}{k_i}$ weakly dominates $b'_i > \frac{v_i}{k_i}$.

If bidder $i$ follows any strategy $b''_i < \frac{v_i}{k_i}$, as long as they are among the winners, the payoff would be the same as the one for a bid equal to $b_i = \frac{v_i}{k_i}$. However, in one situation where the per unit bid of the last winner is between $b''_i$ and $\frac{v_i}{k_i}$ and the knapsack capacity allows $i$ to be packed, $b_i = \frac{v_i}{k_i}$ would result in a positive payoff while $b''_i < \frac{v_i}{k_i}$ results in zero payoff. Therefore $b_i = \frac{v_i}{k_i}$ weakly dominates $b''_i < \frac{v_i}{k_i}$.


Next we prove the uniqueness. First note from Myerson's Lemma for a single-parameter environment we have the following (see \cite{roughgarden2016twenty}):

(A) An allocation rule $\mathbf{x}$ is dominant-strategy incentive-compatible (DSIC) iff it is monotone.

(B) If $\mathbf{x}$ is monotone, then there is a \textit{unique} payment rule such that the sealed-bid mechanism ($\mathbf{x}$, $\mathbf{p}$) is DSIC [assuming the normalization that $b_i = 0$ implies $p_i(b) = 0$]. 

(C) For every bidder $i$, bid $b_i$, and bids $b_{-i}$ by others, the payment rule in (B) is:
$p_i(b_i,b{-i}) = \sum_{j=1}^{l}{z_j} \cdot$ (jump in $x_i$ at $z_j$), where $z_1, z_2, ..., z_l$ are jumps in the allocation function at break points in the range $[0,b_i]$.

For the knapsack auction game the following is true:

(1) The greedy allocation rule is monotonic, since allocation $x_i(z_i,b_{-i})$ is non-decreasing in bid $z_i$. So from (A), it is implementable, that is, there exists a payment rule $\mathbf{p}$ such that $(\mathbf{x},\mathbf{p})$ is DSIC. 

(2) From (B), we know that the DSIC payment rule is unique. 

(3) In (C), since this is a 0-1 allocation, there exists only one critical bid $\hat{z}$ such that for agent $i$: $b_i < \hat{z}$ yields $x_{i} = 0$ and $b_i \geq \hat{z}$ yields $x_{i} = 1$. The question is: what is $\hat{z}$?

\vspace{10pt}

Consider a bid profile $\mathbf{b}$ where $b_{1}>b_{2}>...b_{n}$. Given the Greedy algorithm, suppose that the set of winning bidders be $W=\{1,2,...,m\}$ and the winning bids be $b_{1}>b_{2}>...b{_m}$, where $m<n$. From the Greedy algorithm, this implies that (I) and (II) below are true:

\vspace{5pt}

(I) $\sum_{j \in W}k_{j} \leq K$ and

(II) $\sum_{j \in W}k_{j} + k_{m+1} > K$ 

\vspace{10pt}

Consider an arbitrary winner $i$; the set $W\backslash{i}$ is the winning set of agents excluding $i$. There are two possibilities:

(i) $\sum_{j \in W\backslash{i}}k_{j} + k_{m+1} > K$. In this case the critical bid is trivial: $\hat{z} = b_{m+1}$, as for $b_{i} < b_{m+1}$ we have that $x_i = 0$ due to Greedy algorithm, but for $b_{i} \geq b_{m+1}$, $x_i = 1$ is feasible given (I) above. So the price paid by $i$ equals $b_{m+1}$.

(ii) $\sum_{j \in W\backslash{i}}k_{j} + k_{m+1} \leq K$. Without loss of generality, suppose the removal of $i$ allows agents $\{m+1,m+2,...m+l\}$ to fit their objects in knapsack. In that case, from the Greedy algorithm, we have that $\sum_{j \in W\backslash{i}}k_{j} + k_{m+1} + k_{m+2} + ...k_{m+l} \leq K$, but that $\sum_{j \in W\backslash{i}}k_{j} + k_{m+1} + k_{m+2} + ...k_{m+l} + k_{m+l+1} > K$. Now suppose we start with the lowest bid to size ratio, $b_{m+l}$. This cannot be a critical bid for $i$, because even if $b_{i} > b_{m+l}$, from (II) above we have that $\sum_{j \in W}k_{j} + k_{m+1}>K$, and therefore $i$'s object cannot fit in knapsack. This holds for all bids up to $b_{m+2}$. It is only when $b_{i} \geq b_{m+1}$ that $i$'s object can fit into knapsack and satisfy  the feasibility condition (I). Consequently in this case also $\hat{z} = b_{m+1}$.

Since $i$ was arbitrarily chosen from set $W$, the above holds for all agents in $W$. So, all agents pay $\hat{z} = b_{m+1}$, which implies that the uniform price auction is DSIC.

From Myerson's Lemma, we know that this is unique. So the UP auction is the unique dominant strategy incentive compatible auction.

\end{proof}

The above proposition suggests that UP has the only dominant-strategy and incentive-compatible equilibrium of the knapsack auction game. Upon examining the VCG payment, one can observe situations within the VCG framework where bidders could benefit from overbidding; consequently, we can conclude that VCG is not truthful in this setting.  For instance, consider a scenario where a bidder is the highest losing bidder and the last winning bidder has a low capacity. If the highest losing bidder overbids the last winning bidder, they would only pay more than their per unit value for a small fraction of their capacity, thereby still achieving a positive payoff through overbidding. However, this strategy is not feasible in the UP auction, as the payment for all units would be the same. In this case, it would be equal to the per unit bid of the lowest winning bidder (which will become the highest losing bidder after overbidding), leading to a negative payoff for the overbidding bidder. As we proved the uniqueness of the UP in Proposition \ref{proup}, the following corollary is immediate.

\begin{corollary}
    In the knapsack auction, the VCG payment rule is not truthful. 
\end{corollary}

\begin{proof}
In the proof of proposition \ref{proup} we show the uniqueness of the DSIC and also showed UP is DSIC. What remains to show is that UP is not the same as VCG. It is straightforward to verify that the two payment rules are not the same. Formally denote the value and the size of the highest losing bidder as $v'_i$ and $k'_i$. Also denote the set of winning bidders as $W=\{1,2,...,m\}$. If there is at least one $k_j > k'_i$ for $j\in m$, then the VCG payment rule would become, 

\begin{equation}
    VCG \quad payment =
    \begin{cases}
        & \frac{v'_i}{k'_i} \quad \text{for}  \quad k'_i \quad \text{units}\\
        &\\
        & \frac{v''_i}{k''_i} \quad \text{for} \quad k_j-k'_i \quad \text{units}
    \end{cases}
\end{equation}
where $\frac{v''_i}{k''_i}$ is the per unit value of the second highest loser. Of course we assumed that the sum of the sized of the first and the second highest loser exceeds the size of the winner. Otherwise we move to the consecutive losing bidders until the sum of the sizes reaches the size of the winning bidder. Therefore the VCG payment rule is not necessarily uniform. 
\end{proof}

It is important to note that despite UP having a DSIC equilibrium, it is not necessarily efficient. This can be intuitively understood due to the binary (0-1) nature of packing in the knapsack auction. For instance, recall that the Greedy algorithm stops the allocation when the next object cannot be packed. This essentially means if there is still an object next in the line which has a size less or equal to the remaining knapsack space, it would not be packed. Therefore the truthful nature of the payment rule cannot address the inefficient allocation of the Greedy algorithm. The following proposition formally states the inefficiency of the UP.

\begin{remark}
The UP auction is not efficient.   
\end{remark}

The above remark follows from the fact that since the allocation mechanism is not efficient, even an incentive compatible payment rule cannot achieve efficiency. It suffices to show an example where UP results in an inefficient allocation of the knapsack space. Suppose the lowest winning bidder's capacity is such that there are $\hat{k} > 0$ space left in the knapsack. Denote the lowest winning bidder's per unit value as $\frac{v_i}{k_i}$. Based on the incentive compatible equilibrium, it must be that this per unit value is greater than the highest losing bidder, say $\frac{v'_i}{k'_i}$. If $\hat{k} + k_i \geq k'_i$, then the UP is inefficient as long as $v_i < v'_i$.

As mentioned at the beginning of this section, UP constructs a very good benchmark for our study as it is the only DSIC auction among the three auctions that we investigate. Next we are going to study the equilibrium bidding behavior of the DP auction.
 
\subsection{The Discriminatory Price (DP) auction}

For the DP payment rule it is clear that bidding ones value, that is, bidding $B_i=v_i$, is a weakly dominated strategy as it guarantees zero payoff. However we would like to further investigate if there is any monotone bidding function $\beta(v_i)$ that would characterize a Bayesian-Nash equilibrium of this auction. Each bidder $i$ who submits $B_i$ has the following expected payoff,

\begin{equation}\label{eq1}
\pi_i= (v_i - B_i) Pro(k_i \: \text{being packed})
\end{equation}

To compute the probability of being packed we need to introduce further notation. Denote $K_{-i}$ as the set of all other bidders' capacities excluding bidder $i$. Within this set denote $S$ as a representative subset of $K_{-i}$ with $j$ elements such that there is still space for $i$ to be packed. Formally this can be shown as follows.

\begin{equation}
K - \sum\limits_{j \in S} k_j +k_i < \min_{-j\in K_{-i}\setminus S} k_{-j}
\end{equation}

The probability that bidder $i$ is packed along with all bidders in a subset  $S$ is equal to the probability that all bidder $j$s and bidder $i$ submit the $j+1$ highest per unit bids among all $n$ bidders. Of course this is not the only event that bidder $i$ has a positive probability of being packed and this is only for a given subset $S$. Thus the probability that $k_i$ is being packed is equal to the sum of the probabilities of all those events where it can be packed. 

\begin{equation}
Pro(k_i ; \text{packed}) = \sum_{S \subseteq K_{-i}}  Pro\left(\frac{\beta(v_i)}{k_i} > \max_{t \in K_{-i} \setminus S} \frac{\beta(v_t)}{k_t}\right) \cdot Pro(S)
\end{equation}

Here, $Pro(S)$ is the probability that the objects of the bidders in subset $S$ can be packed along with bidder $i$'s object in the knapsack. This probably is equal to the probability that all bid to size ratios of $j$ members of $S$ is larger than the bid to size ratio of other $t$ members that have not been packed. 

\begin{equation}
    Pro (S) = \sum\limits_{j \in S} Pro\left(\frac{\beta(v_j)}{k_j} > \max_{t \in K_{-i} \setminus S} \frac{\beta(v_t)}{k_t}\right)
\end{equation}

Next we focus on a possible symmetric Bayesian Nash equilibrium $\beta^*$. First note that if such equilibrium exists then it must have two characteristics. First, fixing the size of bidder $i$'s object, a larger value would result in a larger per unit bid. Second, the bid function must be a non-decreasing function of the size, that is, increasing the size of a given bidder's object, their bid $B_i$ should not decline, ceteris paribus. The first point would help us computing the above probabilities using order statistics of the distribution of values. Therefore in a symmetric BNE the probability that $i$'s item is packed with a given subset $S$ is equal to:

\begin{equation}\label{eqprobdp}
Pro(k_i ; \text{packed})\vert_S =  Pro\left(\frac{B_i}{k_i} > \max_{t \in K_{-i} \setminus S} \frac{B_t}{k_t}\right) \cdot \sum\limits_{j \in S} Pro\left(\frac{B_j}{k_j} > \max_{t \in K_{-i} \setminus S} \frac{B_t}{k_t}\right)
\end{equation}

Denote $\psi (v_i, k_i)$ as the probability that bidder $i$ with value $v_i$ and $k_i$ is being packed. The next proposition characterizes the optimal bidding function for this BNE.

\begin{proposition}
   The DP auction has a BNE where each bidder chooses a bid equal to,
   \begin{equation}\label{eqdpbid}
B_i^* = v_i - \frac{\psi(B_i, k_i)}{\frac{\partial \psi(B_i, k_i)}{\partial B_i}}
\end{equation}
\end{proposition}

\begin{proof}
    
Suppose each bidder $i$ would choose their bid $B_i$ to maximize their expected payoff:

\begin{equation}
\max_{B_i} \pi_i = (v_i - B_i) \psi(B_i, k_i)
\end{equation}
Differentiating the above term with respect to $B_i$ gives the following first-order condition:

\begin{equation}
\frac{\partial \pi_i}{\partial B_i} = (v_i - B_i) \frac{\partial \psi(B_i, k_i)}{\partial B_i} - \psi(B_i, k_i)  \ = 0
\end{equation}

This implies:

\begin{equation}
\psi(v_i, k_i) = (v_i - B_i) \frac{\partial \psi(B_i, k_i)}{\partial B_i}
\end{equation}

By rearranging, we can express the optimal bid as an implicit function of firm $i$'s bid, $B_i$, the size $k_i$, and the derivative of the probability of being packed with respect to the bid:

\begin{equation}
B_i^* = v_i - \frac{\psi(B_i, k_i)}{\frac{\partial \psi(B_i, k_i)}{\partial B_i}}
\end{equation}

Also it is routine to check that Equation \ref{eqprobdp} is increasing in $B_i$ and the above is increasing in $v_i$. 
\end{proof}

Based on the above proposition in the BNE of the DP auction each bidder submits a bid that is strictly lower than their value. Note that the $\psi(.,.)$ function is increasing in the bid value $B_i$ as larger bids result in higher probability of being packed, \textit{ceteris paribus}. Therefore the second term of \ref{eqdpbid} is strictly positive. The above result is not surprising as in most of the similar setup the DP auction results in equilibrium with bids strictly below the values and dependent on probability of winning. However, as expounded above the complexity of the knapsack auction is that the probability of winning, which is equivalent to probability of being packed, is very hard to compute due to nature of the problem. 


 \subsection{The Generalized second-price (GSP) auction}

As outlined earier, in the GSP auction format each bidder whose item is packed, pays the per unit bid of the next bidder. Therefore, the expected payoff of a bidder $i$ is as follows.

\begin{equation}\label{eqgsp}
\pi_i= (v_i - k_ib_{i+1}) Pro(k_i \: \text{being packed})
\end{equation}

\begin{remark}
 In the GSP, truth-telling is not an equilibrium.   
\end{remark}





To understand the intuition behind the above remark first note that bidders do not pay their own bid; instead, their payment is computed based on the per-unit bid of the subsequent bidder. Therefore, if a bidder has a high per-unit value, there is an incentive for them to underreport their value to pay less, while ensuring that they remain among the winners. Note that while the above remark suggests that bidding equal to one's value is not an equilibrium in the GSP, it does not comment on the extent of untruthful bidding. According to \cite{Edelman2007}, one could explore a similar type of envy-free equilibrium for the GSP auction that could achieve results akin to those of the VCG mechanism. However, given that the VCG is neither truthful nor efficient in our setup, we refrain from exploring such equilibria. In fact, our primary interest is in understanding, in the absence of such outcomes, what the strategic behavior of bidders in a GSP knapsack auction would be.

\section{Experiment}\label{exp}

Following the theoretical model, we design a laboratory experiment to test the performance of the three proposed auctions. Our first treatment investigates a DP auction where each bidder pays his or her own bid. The second treatment tests the GSP auction where bidders pay based on the bid-to-size ratio of the next bidder. Finally our last treatment tests the UP auction where all bidders pay based on the highest losing bid-to-size ratio. The last treatment not only represents an ideal alternative for a possible knapsack auction but also plays the role of a benchmark as based on the theoretical findings we expect this to be the only strategy proof mechanism among the three. Table \ref{tab2} presents some details regarding the participants in each treatment.

\begin{table}[ht!]
\caption{Treatments summary}
\label{tab2} \centering{}%
\begin{tabular}{l c c c c}
   \toprule
&  Cond. 1 (DP) & Cond. 2 (GSP)  & Cond. 3 (UP) & Total\\[1mm]
\hline\\[-1.5ex]
Participants per group     & 7  & 7  & 7 &  \\[2mm]

No. of groups   & 9  & 8  & 8 & 25\\[2mm]

No. of participants    & 63  & 56  & 56 & 175\\[2mm]
No. of rounds & 20  & 20  & 20 &  \\[2mm]
\hline \\[-1.5ex]
Sum  auction obs. & 180  & 160  & 160& 500 \\[2mm]
Sum  Ind. obs. & 1,260  & 1,120  & 1,120& 3,500\\
\hline
\end{tabular}
\end{table}

Our experiments were run in groups of seven participants who played the auction game for 20 rounds. The knapsack size was fixed to 36 for all the auctions. The size of objects that buyers had were randomly drawn (without replacement) in each round from a set $[4,5,6,7,8,9,10]$, which sums to 49. Therefore, at least two bidders and at most three bidders cannot fit their objects into the knapsack, depending on the bids. Each buyer received an integer value from $[1,10]$, which was distributed uniformly.

\begin{figure}[ht!]
\centering
  \includegraphics[scale=0.39]{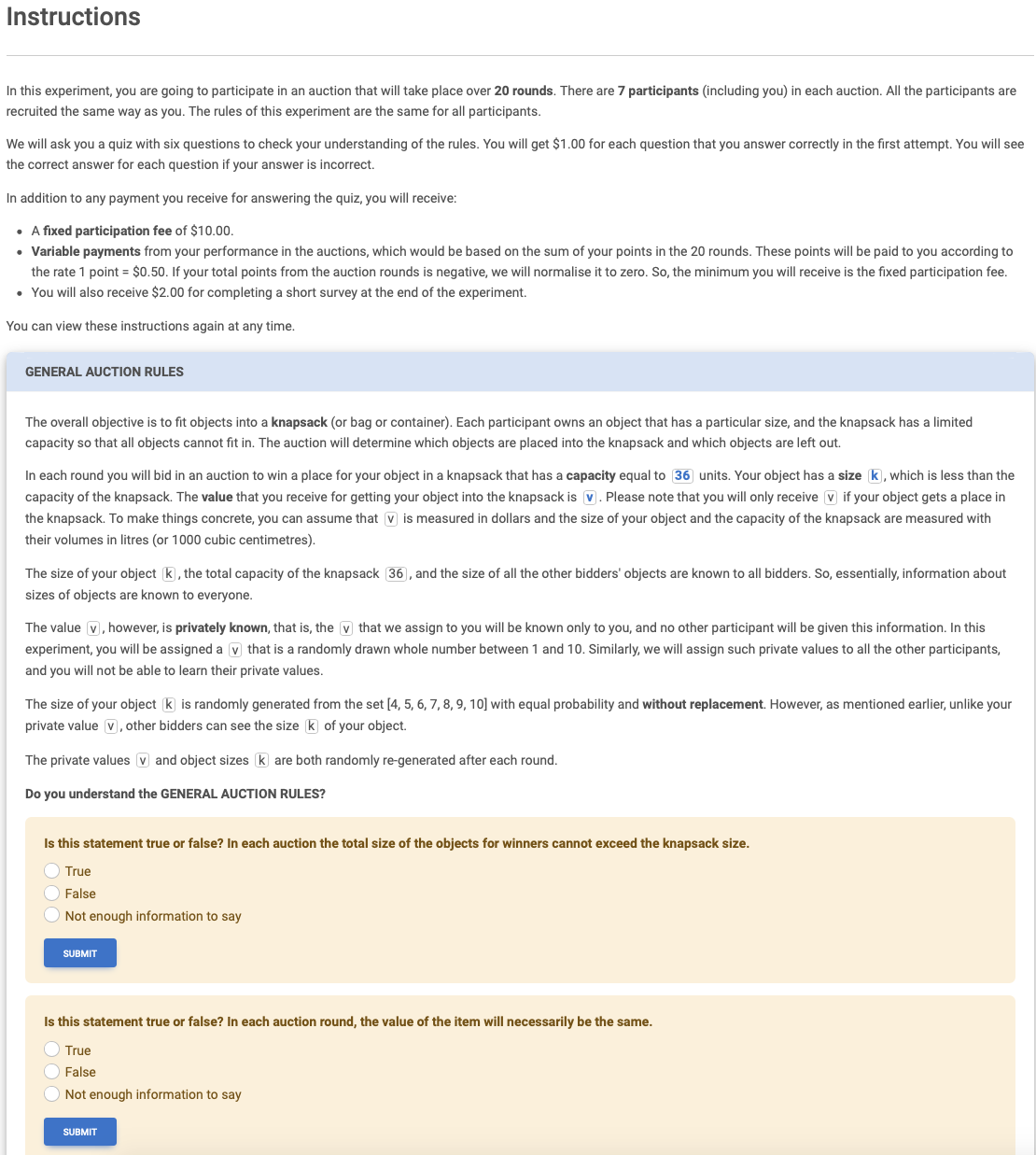}  
\caption{Instructions and quiz questions}
\label{In1}
 \end{figure}

\begin{figure}
\centering
  \includegraphics[scale=0.4]{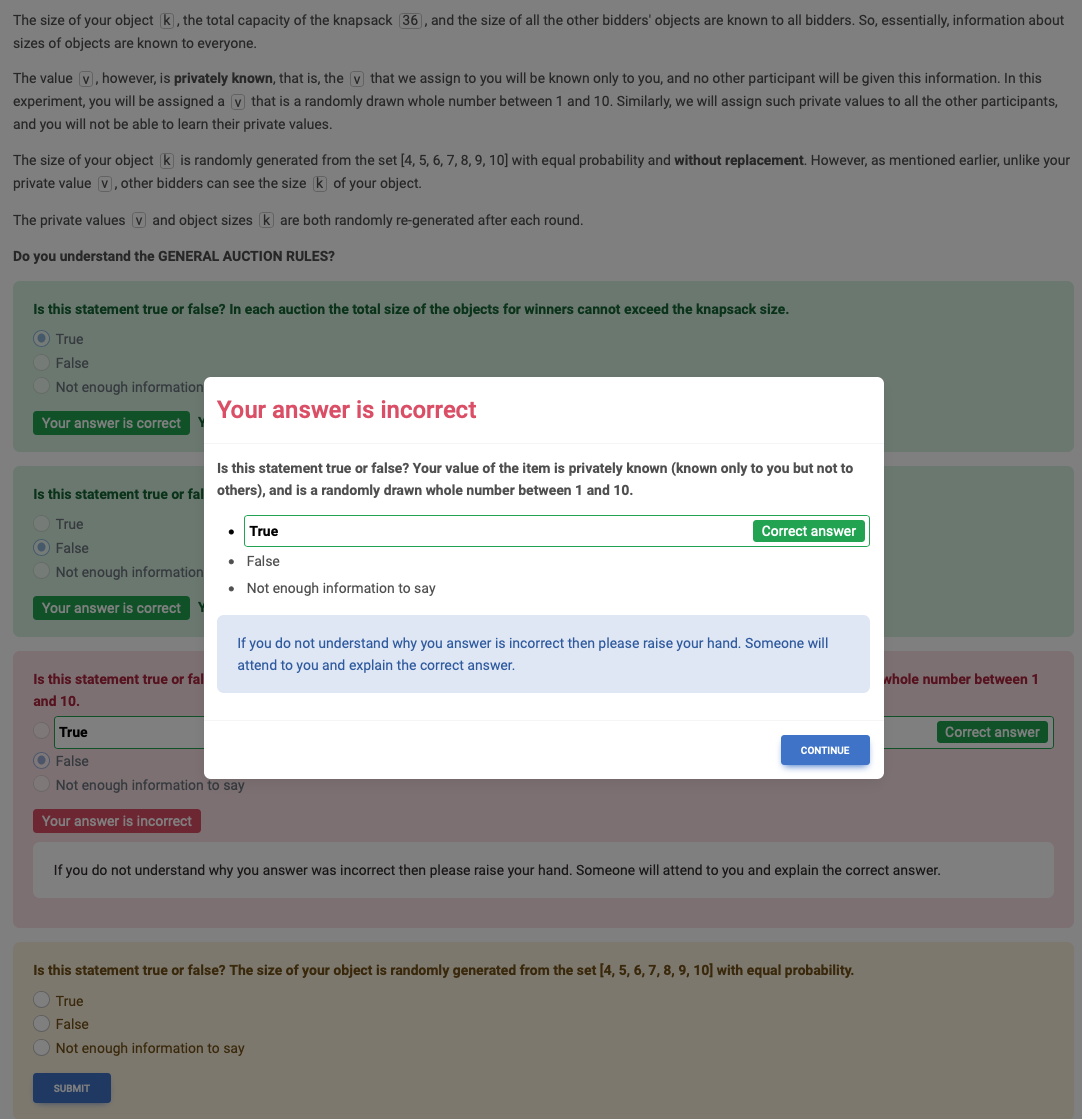}  
\caption{Incorrect quiz questions}
\label{In2}
 \end{figure}

Our experiments were conducted at RMIT's Behavioural Business Lab in Melbourne, Australia. Each session began with subjects being randomly allocated into groups of seven. The duration of each session ranged between 50 to 90 minutes on average, including the time allocated for payments. At the beginning of the session, detailed instructions were provided electronically to the subjects. There were a total of six quiz questions that the subjects needed to answer, which were embedded in the instructions (Figure \ref{In1}). Subjects earned 1 Australian dollar (AUD) for each correct answer. If their answer was incorrect, a pop-up window would appear displaying the correct answer (Figure \ref{In2}). Once they had read the instructions and answered the quizzes, they could proceed to the auction stage. They had the option to access the instructions at any time during the experiment by clicking on a tab at the top of the window.

\begin{figure}
\centering
  \includegraphics[scale=0.39]{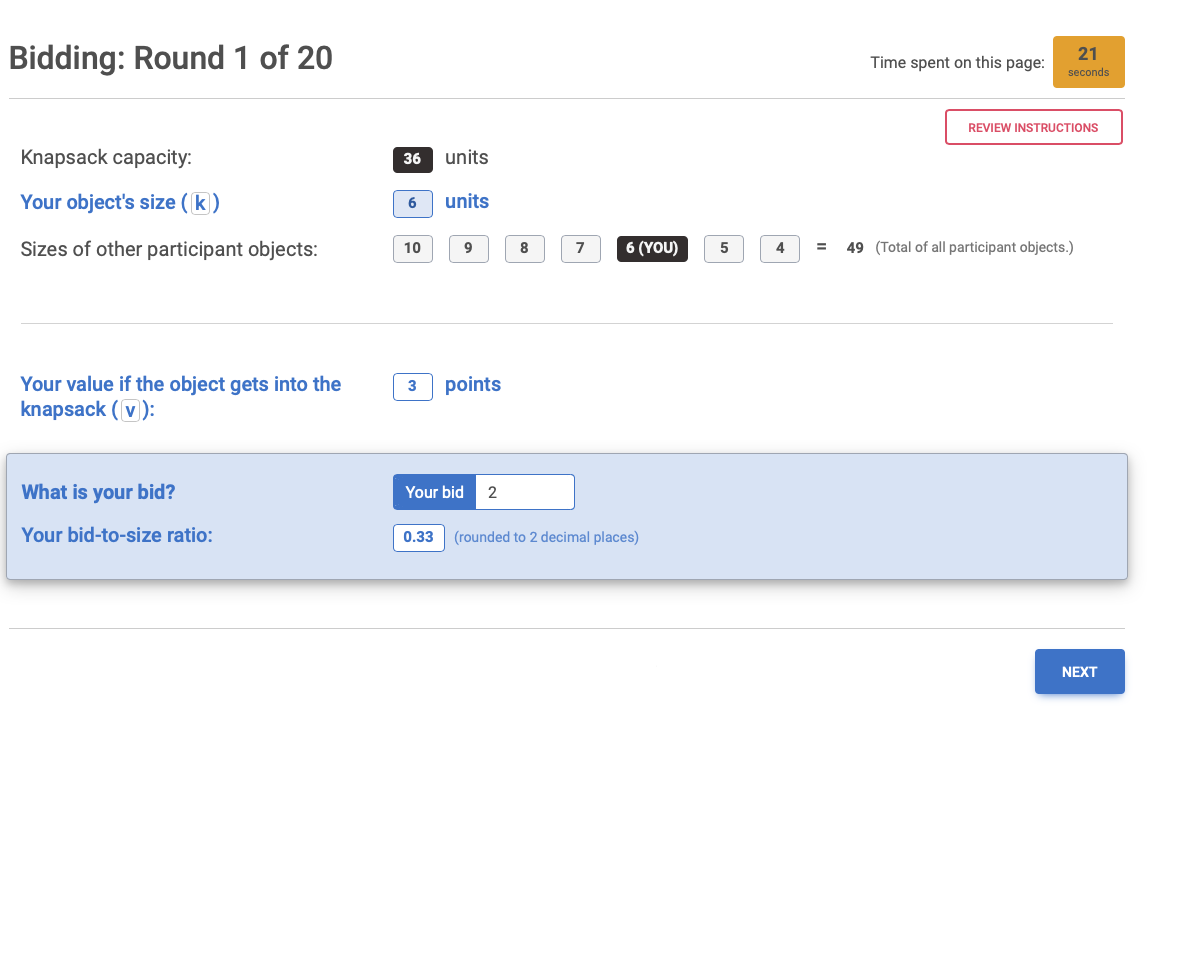}  
\caption{Bidding stage in each round}
\label{Bid1}
 \end{figure}

\begin{figure}
\centering
  \includegraphics[scale=0.39]{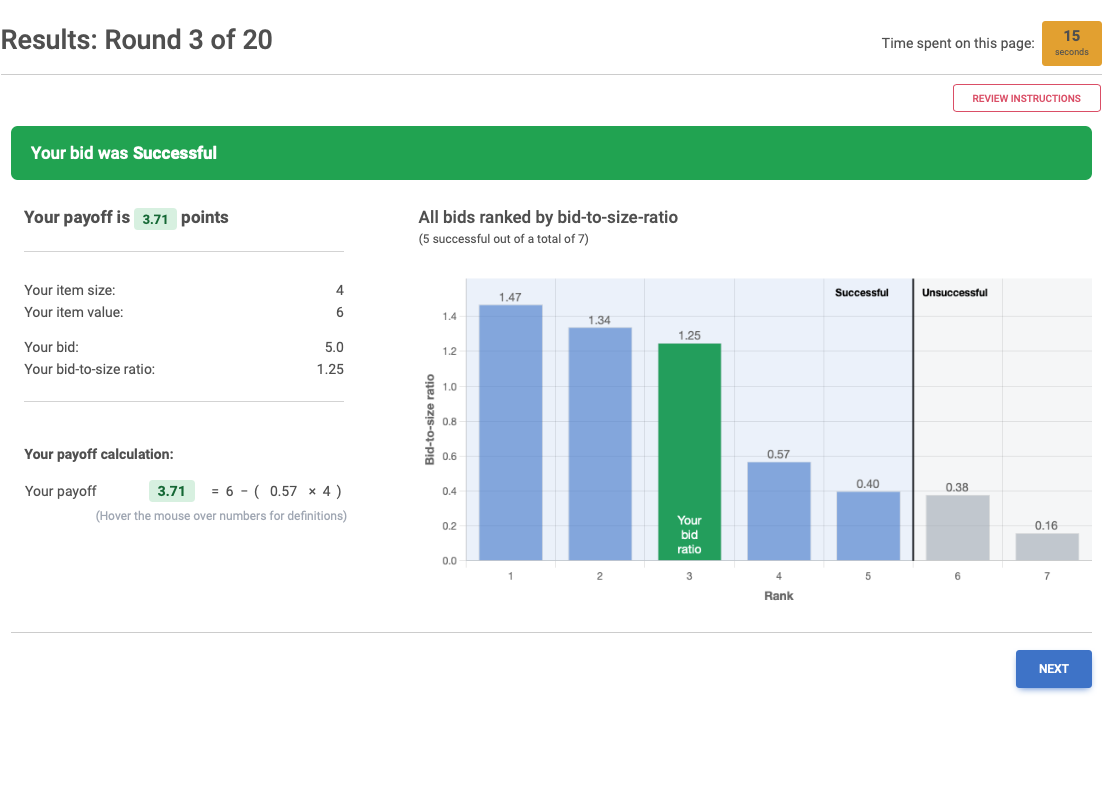}  
\caption{Payoffs in each round. The example is for the GSP treatment.}
\label{Bid2}
 \end{figure}

The auction rounds commenced with each round displaying the knapsack size, the size of the object owned by the bidder, and the size of the objects owned by other bidders, including the sum (Figure \ref{Bid1}). Subsequently, participants were required to place their bids for the item. Participants could view their bid-to-size ratio below their bid entries, which ultimately determines their winning status. After all subjects submitted their bids, they were redirected to a subsequent page. This page revealed whether they emerged as winners or not, their respective payoffs, the calculation process for these payoffs, and the bids submitted by other bidders (Figure \ref{Bid2}).

Subjects accrued experimental points during the auction stage, which were converted to dollars at an exchange rate of 2 points to 1 AUD. Their earnings from the auction stage was the sum of their payoffs from each round. In the event of a negative payoff in a given round, this amount was deducted from the total. Subjects were informed that if the cumulative earnings from the auction stage were negative, it would be normalized to zero. However, no instances of total negative earnings were recorded. In addition to this, subjects also received a 10 AUD participation fee and 2 AUD for completing a post-experiment survey, which included two questions evaluating their risk preferences. On average, subjects earned 35.24 AUD (approximately USD 24.66), with a range from a minimum of 16 to a maximum of 52.50 AUD. 

\subsection{Experimental results}

There are three major variables of interest in each auction: bids placed by bidders, revenue and efficiency of the auction. To be able to to provide a comprehensive analysis of each auction we have introduced three measurements for each of these variables as follows.

\textbf{Learning ratio} is the difference between per unit value and per unit bid. Formally one can define learning ratio as,

\[ R= \frac{v_i}{k_i} - \frac{B_i}{k_i}\]

This variable aims to capture how close bids are to the value of the bidder. We use the same variable in the next section to measure the learning patterns of AI agents. 

\textbf{Revenue} is the sum of payments by all those who had successful bids. In case of DP this is simply the sum of all the winning bids. In case of GSP it is the sum of the payments of all successful bidders, where every winner pays the bid-to-size ratio submitted by the next highest bidder times their item size. And for the UP is the bid to size ratio of the highest losing bidder times the size of every winner. As we already have this information to compute the payoff of each player, we only need to sum the payments for all the successful bidders to compute the revenue of each auction.
 
\textbf{Efficiency (surplus)}: To compute the efficiency, we first need to calculate the value-to-size ratio for all bidders in each auction. Then, starting from the highest value-to-size ratio, we fill the knapsack until it is full. We then compute the sum of the total values of those who were packed (denote this sum as $S$). This sum represents the solution given by the Greedy algorithm in the case of full information. The next step is to compute the sum of the values of those who are packed based on the bidding in the auction, that is, the sum of the total values of successful bidders (denote this sum as $C$). The efficiency of each auction is represented by the magnitude of 
$S-C$. The closer $C$ is to $S$, the more efficient the auction is. For instance, for the UP auction, we expect to have a very small (or zero) $S-C$ as bidders are incentivized to truthfully reveal their values.

Our main goal is to test how the three auctions perform in terms of truthfulness of the bids, revenue and efficiency. 

Our secondary hypothesis: 
\begin{itemize}
    \item In the UP auction bidders bid truthfully, that is, their bids are equal to their value.
   \item  In the GSP auctions bidder submit untruthful bids, that is, bids are strictly below their value. 
    \item The magnitude of Bidders' bids are ranked as follows: UP results in the highest bids, then GSP and then DP in the lowest bids submitted by bidders. (note this is irrespective of the revenue generated by each auction) 
\end{itemize}

\begin{figure}[ht!]
\centering
  \includegraphics[scale=0.47]{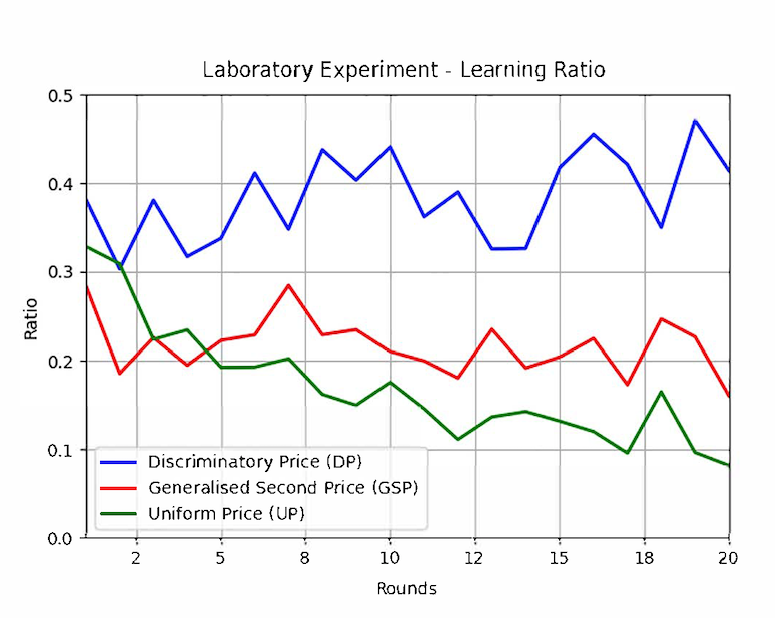}
\caption{Mean individual observations in a round.}
\label{expLR}
 \end{figure}

Figure \ref{expLR} depicts the performance of the three auctions regarding the bids. We used the learning ratio $R= \frac{v_i}{k_i} - \frac{B_i}{k_i}$ described above to represent the truthfulness of bids in the auction. Obviously the smaller the $R$ the more truthful the auction. As shown in Figure \ref{expLR} UP results in the most truthful bids followed by GSP and DP.

\begin{figure}[ht!]
\centering
  \includegraphics[scale=0.8]{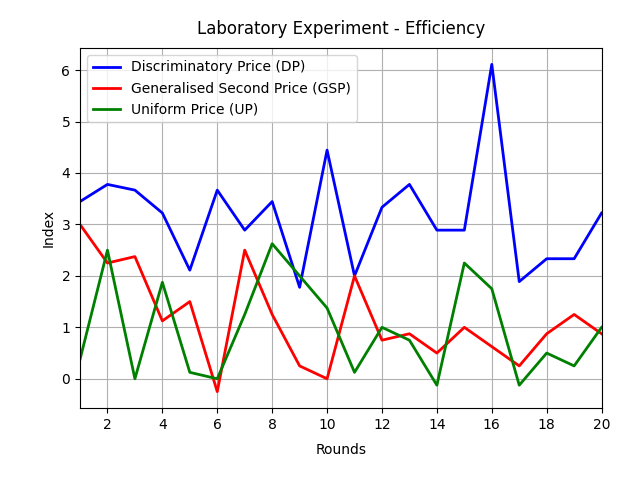}
\caption{7 players in each auction, played for 20 rounds.}
\label{expeff}
 \end{figure}

Next we investigate the efficiency of each auction, Efficiency, as described before is formulated as $E= S- C $ where $S$ is the achievable surplus with full information Greedy algorithm and, $C$ is the achieved surplus via the implemented auction. As shown in Figure \ref{expeff} UP and GSP result in the highest level of efficiency follow  by DP. In some instances the $E$ was zero, indicating that the auctions have allocated the objects fully efficiently.

\begin{figure}[ht!]
\centering
  \includegraphics[scale=0.8]{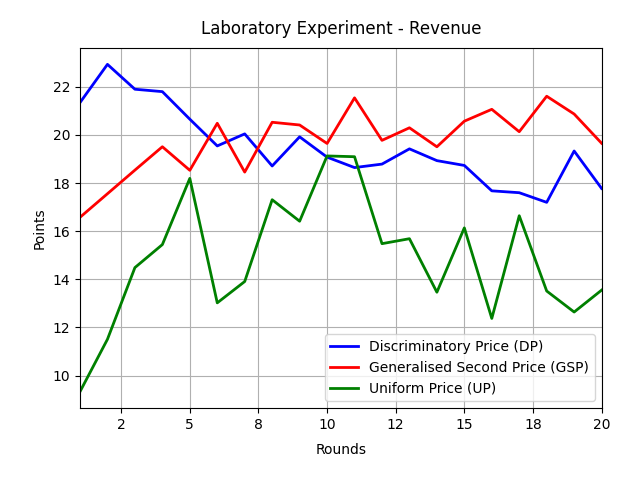}
\caption{7 players in each auction, played for 20 rounds.}
\label{exprev}
 \end{figure}

Finally we compare the revenue obtained by each of the three auctions. As shown in Figure \ref{exprev} GSP resulted in the highest level of revenue followed by DP and UP. While the revenue generated by GSP is very close to the DP, but UP resulted in a significantly lower revenue compared to the other two auctions.

\subsection{Regression results}

In this subsection we present the analysis of the experimental data. First, we start by the linear regression of the three conditions, considering DP as the control. We run two regressions one for the revenue and one for the efficiency of each auction. For the analysis of bidding behavior we used individual data for each of the auctions. However for the revenue and efficiency, since the analysis is at auction level, the data is clustered to 25 clusters where each of these cluster represents a given group of participant observed over 20 rounds of auction.

\begin{table}[ht!]
    \centering
    \begin{tabular}{l c c c c}
    \hline
        Learning ratio & Coef. & Robust SE & t & p-value \\
        \hline
        
  Condition       &  &  &  & \\
  \hline
      GSP   & -0.152 & 0.036 & -4.17 & 0.000 \\
       UP  & -0.209 &  0.035 & -5.95 & 0.000\\
        Cons & 0.344 & 0.026 & 6.23 & 0.000\\
        \hline
        
        Number of obs       & 3,482 &  &  & \\
        R-squared      & 0.051 &  &  & \\
        \hline
    \end{tabular}
    \caption{Linear regression: Learning ratio. Data: total number of individual observations excluding $R<-2$. Note: Excluded 18 observations where some bidders significantly overbid in the UP auction.}
    \label{lreg0}
\end{table}

Table \ref{lreg0} shows the result of a linear regression for the learning ratio of the three auctions, where DP is the dependent variable. As the results show, GSP and UP have significantly higher truthful bids compared to DP. 

Table \ref{lreg1} and \ref{lreg2} show the results for the linear regressions where the round efficiency and revenue for DP are the dependent variables.

\begin{table}[ht!]
    \centering
    \begin{tabular}{l c c c c}
    \hline
        Round Efficiency & Coef. & Robust SE & t & p-value \\
        \hline
        
  Condition       &  &  &  & \\
  \hline
      GSP   & -2.011 & 0.536 & -3.75 & 0.001 \\
       UP  & -2.186 &  0.520 & -4.20 & 0.000\\
        Cons & 3.161 & .507 & 6.23 & 0.000\\
        \hline
        
        Number of obs       & 500 &  &  & \\
        R-squared      & 0.108 &  &  & \\
        \hline
    \end{tabular}
    \caption{Linear regression: Efficiency. Data: 500 auctions (25 groups participating in 20 rounds of auctions). Standard errors clustered by group.}
    \label{lreg1}
\end{table}

As shown in Table \ref{lreg1} both GSP and UP have significantly higher levels of efficiency compared to the UP. Negative coefficients for efficiency essentially means lower index level $E$ for those two auctions which is equivalent to a higher surplus in each auction. Comparing the coefficients of the GSP and UP shows that UP results in higher efficiency compared to GSP, however for a very small and negligible amount. 

\begin{table}[ht!]
    \centering
    \begin{tabular}{l c c c c}
    \hline
        Round Revenue & Coef. & Robust SE & t & p-value \\
        \hline
        
  Condition       &  &  &  & \\
  \hline
      GSP   & 0.261 & 0.888 & 0.29 & 0.771 \\
       UP  & -4.634 &  1.084 & -4.27 & 0.000\\
        Cons & 19.502 & .788 & 24.75 & 0.000\\
        \hline
        
        Number of obs       & 500 &  &  & \\
        R-squared      & 0.206 &  &  & \\
        \hline
    \end{tabular}
    \caption{Linear regression: Revenue. Data: 500 auctions (25 groups participating in 20 rounds of auctions). Standard errors clustered by group.}
    \label{lreg2}
\end{table}

 Table \ref{lreg2} shows the result of regressions for the revenue of each auction. As shown in this table, UP  has a significantly lower revenue compared to the DP. However, the difference between the revenue generated by GSP and UP is not significant. When we exclude the first 10 rounds and only consider the second 10 rounds then the coefficient for GPS becomes positive and significant indicating that GSP results in a higher revenue than DP when subjects completed 10 rounds and learned more about the bidding (Table \ref{lreg3}). This result demonstrates that GSP has the overall highest revenue and efficiency between the three auctions.

 \begin{table}[ht!]
    \centering
    \begin{tabular}{l c c c c}
    \hline
        Round Revenue & Coef. & Robust SE & t & p-value \\
        \hline
        
  Condition       &  &  &  & \\
  \hline
      GSP   & 2.092 & 0.818 & 2.56 & 0.017 \\
       UP  & -3.550 &  0.969 & -3.66 & 0.001\\
        Cons & 18.41 & 0.615 & 29.93 & 0.000\\
        \hline
        
        Number of obs       & 250 &  &  & \\
        R-squared      & 0.238 &  &  & \\
        \hline
    \end{tabular}
    \caption{Linear regression: Revenue. Data: 250 auctions (25 groups for the last 10 rounds of auctions). Standard errors clustered by group.}
    \label{lreg3}
\end{table}



\section{AI Simulations}\label{AI}

In this section, we use simulation methods featuring artificially intelligent agents who bid and learn in the aforementioned three auctions. The aim is to complement the outcomes from these simulations with those from the experiments and to investigate how simulation outcomes might address the shortcomings of the experiments. For our simulations, we used a Q-learning algorithm to enable agents to learn in an environment that closely resembles our experimental setup.

Q-learning algorithms have increasingly been utilized to simulate human behavior in auction environments, offering insights into complex decision-making processes \citep{banchio2022artificial}. These algorithms, rooted in the field of reinforcement learning, are adept at handling situations where agents interact under uncertainty and incomplete information – conditions typical of auction markets. By iteratively updating their strategy based on rewards or punishments received from the environment (in this case, the auction outcomes), Q-learning agents can learn how to bid optimally in auctions.\footnote{\cite{khezr2024artificial} is a recent study that summarizes various reinforcement learning algorithms. It suggests that, despite its simplicity, Q-learning performs relatively well compared to more advanced reinforcement learning algorithms in simulating bidding in multi-unit auctions.} They adapt their bidding strategies based on past experiences and perceived patterns, reflecting the dynamic nature of human decision-making. This approach allows us to explore various auction formats and bidding behaviors, providing valuable perspectives on how individuals might act in competitive bidding scenarios.

Q-learning is a model-free, off-policy reinforcement learning approach. Utilizing the widely adopted Q-Learning algorithm, our method is designed to learn a bidding strategy through a sequence of episodes (auctions), each comprising a single auction with multiple agents (bidders). Learning takes place in a decentralized environment, where each agent learns independently, relying solely on observations from their private states, actions, and rewards. A state ($s$) represents an agent's current auction `environment', including the value and size of their item, assigned at the start of each episode. An action ($a$) is the agent's bid, determined by their current state, aiming to secure their item in the knapsack while maximizing their payoff. The reward ($r$) is derived from the combination of state and action $<s, a>$. Agents whose actions result in their item being included in the knapsack receive their payoff as a reward. Conversely, if an agent's action does not secure a place in the knapsack for their item, they receive a negative reward.

Over a series of $n$ episodes, each agent populates a Q-table, recording rewards from various state/action pairs. Initially, actions are taken randomly over several episodes to start populating the table and to explore the state/action space. Actions taken during this phase are `exploratory', while `exploitative' actions later involve selecting bids from the agent's Q-table that yield the highest reward given the current state. After the initial exploratory period, a monotonically decreasing `exploration rate' governs the balance between `exploration' and `exploitation', culminating in pure exploitation at the end of the $n$ episodes.

An agent’s Q-table is updated at the end of each episode $e_1$, based on the reward $r_1$ observed from taking action $a_1$ in state $s_1$. The update uses a modified Bellman Optimality Equation, excluding the discount factor ($\gamma$). The discount factor typically balances immediate versus future rewards in a learning task. However, as each auction in our simulations is independent, comprising only a single state/action decision per agent, the discount factor is irrelevant in this context.

\begin{figure}[ht!]
\centering
  \includegraphics[scale=0.8]{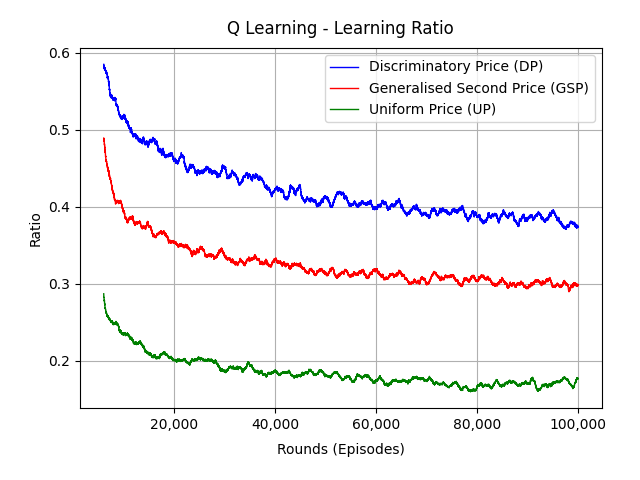}
\caption{7 agents in each auction, played for 100,000 episodes.}
\label{AILR}
 \end{figure}

In our simulations seven Q-learning agents play each auction for 100,000 episodes. All other parameters such as the distributions and the knapsack size are similar to the experimental parameters. The first 1,000 rounds were pure learning and exploitation started afterwards.  

Figure \ref{AILR} shows the results of the learning ratio as defined in the previous section. The importance of the learning ratio is mostly obvious in this figure as one expects if agents learn properly the learning ration must not have a high variance after a given episode. Based on Figure \ref{AILR} there is a high and in some sense natural rate of fluctuation in the learning ratio in the first 40,000 episodes. In these episodes agents are exploring to learn their optimal strategy. However once we pass the 50,000 episodes the change in the learning ratio becomes very small which indicates that the agents have now learned enough about their optimal strategy in the auction.

As shown in Figure \ref{AILR} the DP auction results in the highest ratio followed by GSP and UP. This indicates that in the UP auction Q-learning agents' bids were closest to their values compared to the other two auctions. 

\begin{figure}[ht!]
\centering
  \includegraphics[scale=0.8]{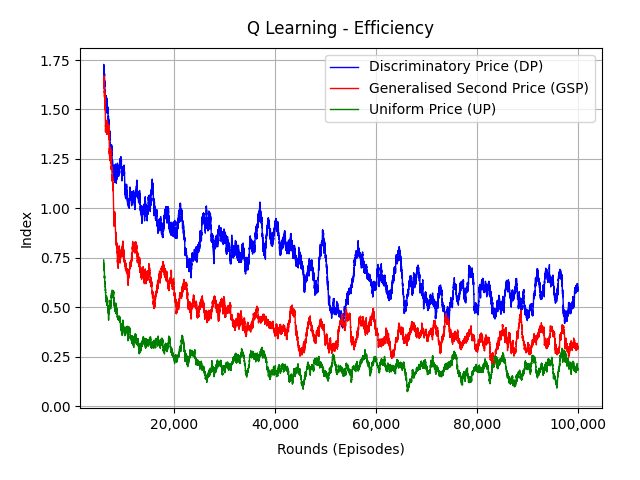}
\caption{7 agents in each auction, played for 100,000 episodes.}
\label{AIeff}
 \end{figure}

Figure \ref{AIeff} shows the efficiency levels of each auction in the Q-learning simulations. As demonstrated, UP exhibits the highest level of efficiency, followed closely by GSP, while DP has the lowest efficiency among the three auctions.

\begin{figure}[ht!]
\centering
  \includegraphics[scale=0.8]{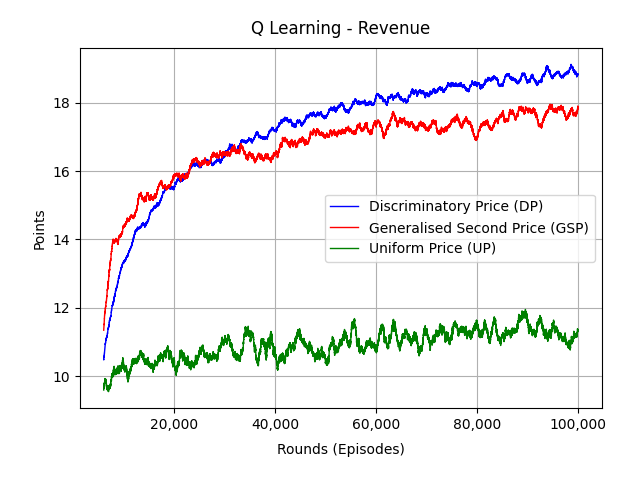}
\caption{7 agents in each auction, played for 100,000 episodes.}
\label{AIrev}
 \end{figure}

Finally, Figure \ref{AIrev} displays the revenue generated in each auction across all episodes. As illustrated in this figure, DP yields the highest revenue level, followed closely by GSP. However, UP generates significantly lower revenue compared to the other two auctions. While similar results were observed in the lab experiment, the difference in revenue between UP and the other auctions is much more pronounced in the AI simulations relative to the experimental data.

One advantage of AI simulations over lab experiments is that, once a suitable algorithm is developed, it can easily run various scenarios with different parameters. In Appendix \ref{appcomp}, we present several comparative statics analyses, altering all parameters to examine the robustness of our results to these changes. Overall, the results demonstrate minimal to no changes in terms of revenue ranking and efficiency when parameters such as knapsack size and the number of agents are varied. This indicates that our findings are robust and can be extrapolated beyond just the parameters employed in the experiments.

\section{Conclusions}

In this paper, we study the well-known knapsack problem in a setting where the knapsack owner and object owners are distinct parties. Object owners possess private information about the value they derive if their object is packed in the knapsack. Focusing on the Greedy algorithm as the allocation mechanism, we examine three auctions for the payment rules governing the transfer between the knapsack owner and the object owners. Two of these auctions, namely the discriminatory price and the generalized second-price auctions, have been widely used in practice for various applications. The uniform price auction was chosen as a benchmark, as we demonstrate it to be the unique truthful mechanism for the knapsack auction problem with incomplete information.

We used three distinct methodologies to compare the three auctions. Starting with a theoretical model, we demonstrate that the DP auction has a Bayesian Nash Equilibrium (BNE) that is challenging to compute, even with simple parameters. We then show that the GSP auction is not truthful, even though bidders do not pay their own bid. Finally, we establish that the UP auction is incentive compatible but inefficient, as the allocation mechanism itself is not efficient.

We also conduct a series of lab experiments with human subjects to test the performance of these auctions. Based on the lab results, the UP auction is the most truthful and efficient among the three, yet it generates the lowest revenue. The DP auction performs well in terms of revenue but is significantly less efficient compared to the other two. The GSP auction closely matches the UP auction in terms of efficiency and resembles the DP auction in revenue, indicating it may be one of the best auctions to use in practice.

Finally, we utilize AI simulations where our agents are trained using a Q-learning algorithm. The AI results mostly align with the lab experiments, with one major difference: the UP auction's revenue is significantly lower than the other two auctions in the AI simulations, in contrast to the lab experiment where the difference, though still notable, is less pronounced.



\clearpage
\section{Appendix: Comparative statics}\label{appcomp}

\begin{table}[ht!]
    \def\arraystretch{1.2}
    \captionsetup{width=1\textwidth}
    \setlength\tabcolsep{7.4pt}
    \caption{\textbf{Agent learning ratios: Randomly selected run (100,000 rounds)}\label{tbl:agent_learning_ratio}}
    \centering
    \footnotesize
    \begin{tabular}{l|rrr|rrr|rrr}
        \toprule
        Auction & \multicolumn{3}{c|}{\textbf{All agents}} & \multicolumn{3}{c|}{\textbf{Worst performing agent}} & \multicolumn{3}{c}{\textbf{Best performing agent}} \\
        type & \multicolumn{1}{r}{Median} & \multicolumn{1}{r}{Mean} & \multicolumn{1}{r|}{SD} & \multicolumn{1}{r}{Median} & \multicolumn{1}{r}{Mean} & \multicolumn{1}{r|}{SD} & \multicolumn{1}{r}{Median} & \multicolumn{1}{r}{Mean} & \multicolumn{1}{r}{SD} \\
        \midrule
        \textbf{DP} & 0.288 & 0.421 & 0.409 & 0.165 & 0.410 & 0.393 & 0.166 & 0.433 & 0.427 \\
        \textbf{GSP} & 0.206 & 0.325 & 0.337 & \textbf{0.163} & 0.323 & 0.336 & \textbf{0.165} & 0.337 & 0.338 \\
        \textbf{UP} & \textbf{0.107} & \textbf{0.185} & \textbf{0.238} & 0.164 & \textbf{0.205} & \textbf{0.281} & \textbf{0.165} & \textbf{0.171} & \textbf{0.249}\\
        \bottomrule
    \end{tabular}
\end{table}

\begin{table}[ht!]
    \def\arraystretch{1.2}
    \captionsetup{width=1\textwidth}
    \setlength\tabcolsep{7.4pt}
    \caption{\textbf{Agent payoff (points): Randomly selected run (100,000 rounds)}\label{tbl:agent_payoff}}
    \centering
    \footnotesize
    \begin{tabular}{l|rrr|rrr|rrr}
        \toprule
        \textbf{Auction} & \multicolumn{3}{c|}{\textbf{All agents}} & \multicolumn{3}{c|}{\textbf{Worst performing agent}} & \multicolumn{3}{c}{\textbf{Best performing agent}} \\
        \textbf{type} & \multicolumn{1}{r}{Median} & \multicolumn{1}{r}{Mean} & \multicolumn{1}{r|}{SD} & \multicolumn{1}{r}{Median} & \multicolumn{1}{r}{Mean} & \multicolumn{1}{r|}{SD} & \multicolumn{1}{r}{Median} & \multicolumn{1}{r}{Mean} & \multicolumn{1}{r}{SD} \\
        \midrule
       \textbf{DP} & 1.520 & 2.113 & 2.169 & \textbf{1.710} & 2.052 & \textbf{2.108} & 1.708 & 2.183 & 2.235 \\
        \textbf{GSP} & 1.800 & 2.219 & \textbf{2.154} & \textbf{1.707} & 2.194 & 2.161 & \textbf{1.710} & 2.254 & \textbf{2.158} \\
        \textbf{UP} & \textbf{2.704} & \textbf{3.087} & \textbf{2.854} & 1.700 & \textbf{3.065} & \textbf{2.860} & \textbf{1.710} & \textbf{3.110} & \textbf{2.855}\\
        \bottomrule
    \end{tabular}
\end{table}

\begin{table}[ht!]
    \def\arraystretch{1.2}
    \captionsetup{width=1\textwidth}
    \setlength\tabcolsep{15.4pt}
    \caption{\textbf{Auction performance: Randomly selected run (100,000   rounds)}\label{tbl:auction_performance}}
    \centering
    \footnotesize
    \begin{tabular}{l|rrr|rrr}
        \toprule
        \textbf{Auction} & \multicolumn{3}{c|}{\textbf{Revenue (points)}} & \multicolumn{3}{c}{\textbf{Efficiency (ratio)}} \\
        \textbf{type} & \multicolumn{1}{r}{\textbf{Median}} & \multicolumn{1}{r}{\textbf{Mean}} & \multicolumn{1}{r|}{\textbf{SD}} & \multicolumn{1}{r}{\textbf{Median}} & \multicolumn{1}{r}{\textbf{Mean}} & \multicolumn{1}{r}{\textbf{SD}} \\
        \midrule
        \textbf{DP} & \textbf{17.400} & \textbf{17.280} & \textbf{3.055} & \textbf{100} & 97.959 & \textbf{5.617} \\
        \textbf{GSP} & 16.643 & 16.766 & \textbf{3.760} & \textbf{100} & 98.744 & 4.484 \\
        \textbf{UP} & \textbf{9.581} & \textbf{10.923} & \textbf{6.418} & 100 & \textbf{99.337} & \textbf{3.480}\\
        \bottomrule
    \end{tabular}
\end{table}

\clearpage

\begin{figure*}[ht!]
    \captionsetup[sub]{font=scriptsize,labelfont={bf,sf}}
    \centering
    \begin{subfigure}[t]{.48\textwidth}
    \includegraphics[width=1.0\linewidth]{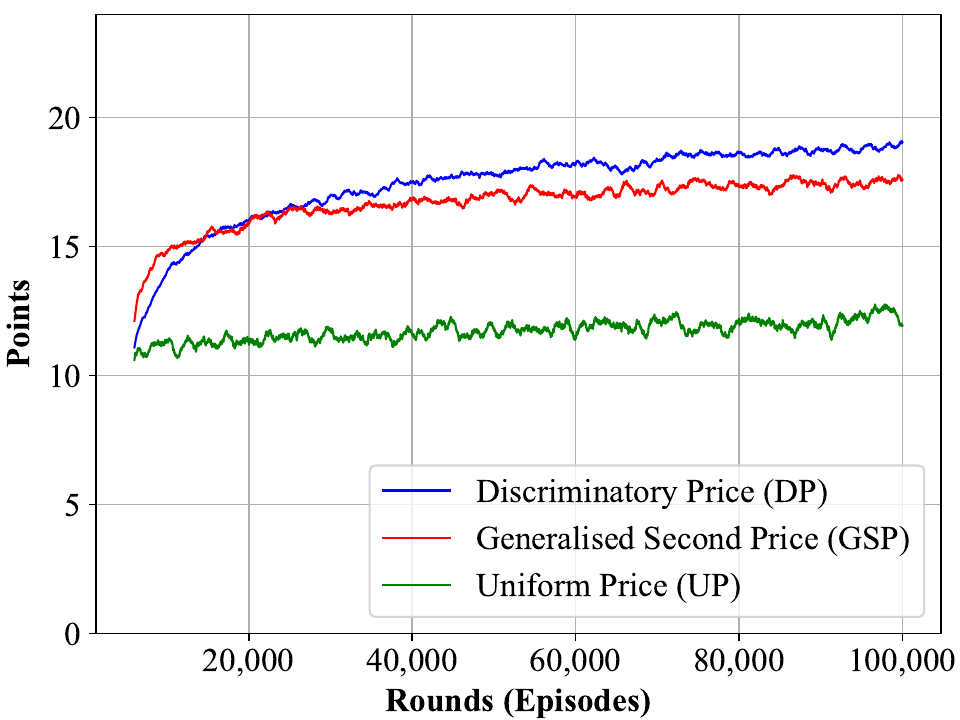}
    \caption{Auction revenue\label{fig:Revenue_7_30_1-10_4-10}}
    \end{subfigure}%
     \begin{subfigure}[t]{.48\textwidth}
    \includegraphics[width=1.0\linewidth]{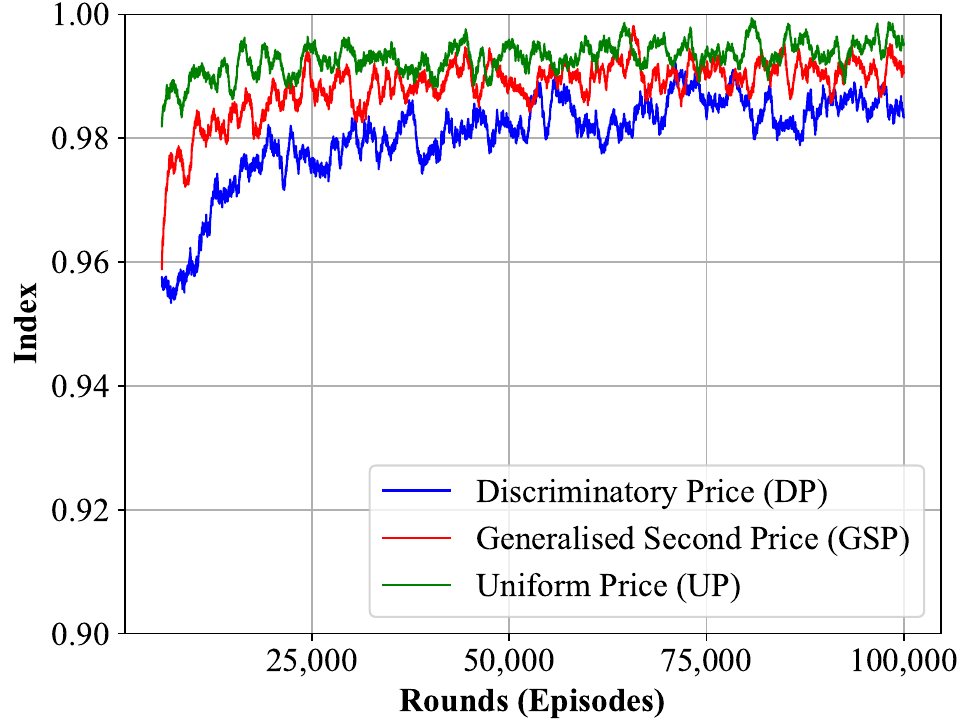}
    \caption{Auction efficiency\label{fig:Efficiency_7_30_1-10_4-10}}
    \end{subfigure}%
    \hfill\vspace{2mm}
    \begin{subfigure}[t]{.48\textwidth}
    \includegraphics[width=1.0\linewidth]{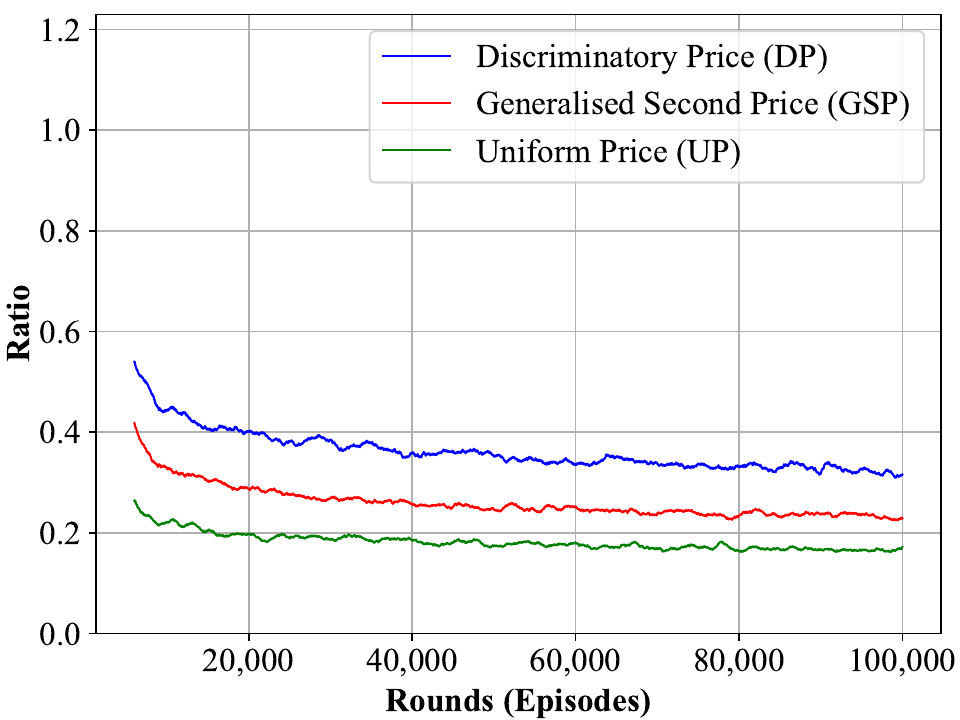}
    \caption{\centering Learning ratio: Average of 7 Q-learning agents\label{fig:Agents_7_30_1-10_4-10}}
    \end{subfigure}%
    \caption{7 Agents, Knapsack capacity of 30, item values in the range [1,10] and item sizes in the range of [4,10]\label{fig:All_7_30_1-10_4-10}}
\end{figure*}

\begin{figure*}[ht!]
    \captionsetup[sub]{font=scriptsize,labelfont={bf,sf}}
    \centering
    \begin{subfigure}[t]{.48\textwidth}
    \includegraphics[width=1.0\linewidth]{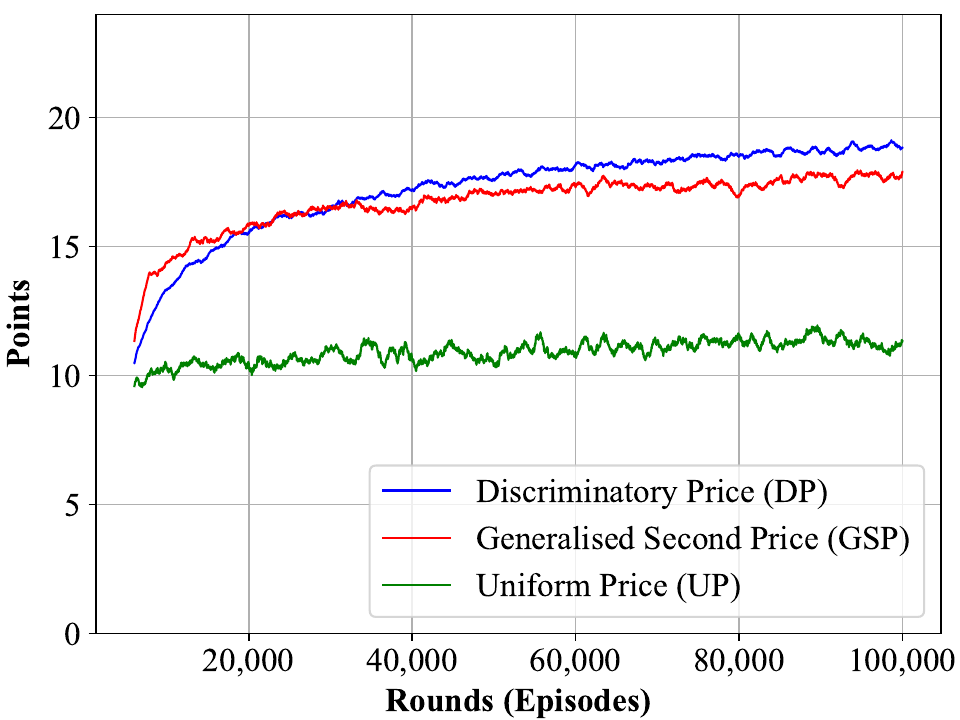}
    \caption{Auction revenue\label{fig:Revenue_7_36_1-10_4-10}}
    \end{subfigure}%
     \begin{subfigure}[t]{.48\textwidth}
    \includegraphics[width=1.0\linewidth]{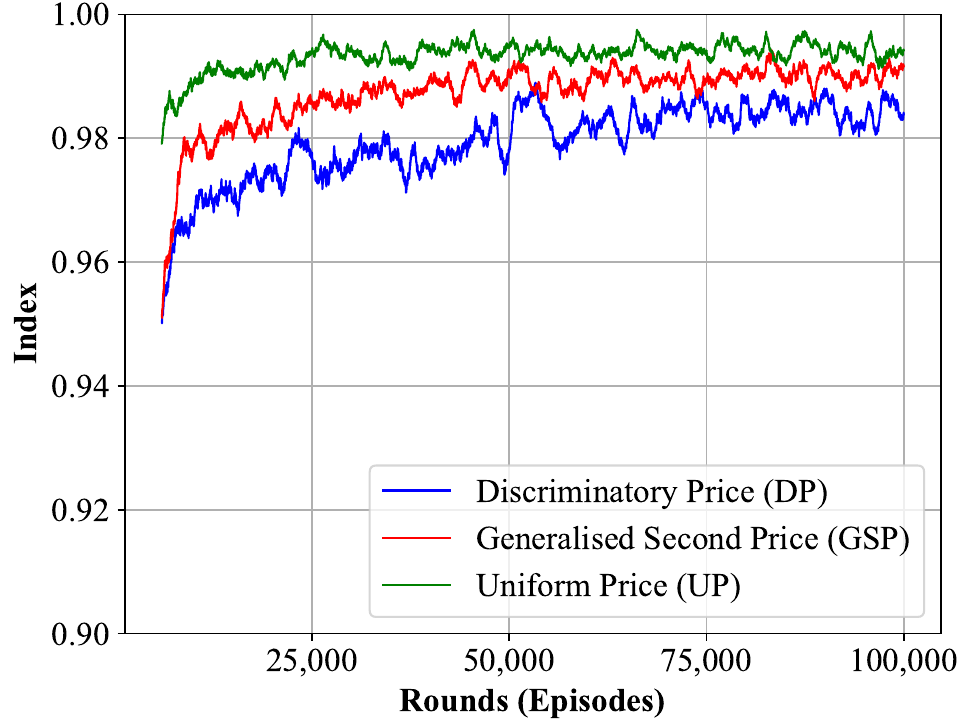}
    \caption{Auction efficiency\label{fig:Efficiency_7_36_1-10_4-10}}
    \end{subfigure}%
    \hfill\vspace{2mm}
    \begin{subfigure}[t]{.48\textwidth}
    \includegraphics[width=1.0\linewidth]{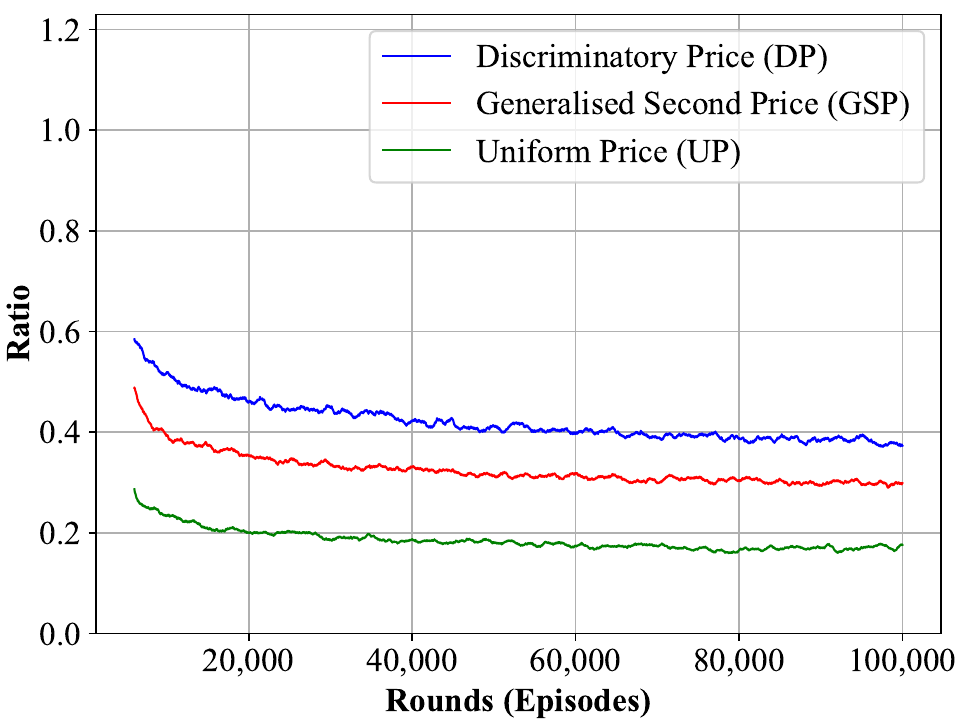}
    \caption{\centering Learning ratio: Average of 7 Q-learning agents\label{fig:Agents_7_36_1-10_4-10}}
    \end{subfigure}%
    \caption{7 Agents, Knapsack capacity of 36, item values in the range [1,10] and item sizes in the range of [4,10]\label{fig:All_7_36_1-10_4-10}}
\end{figure*}

\begin{figure*}[ht!]
    \captionsetup[sub]{font=scriptsize,labelfont={bf,sf}}
    \centering
    \begin{subfigure}[t]{.48\textwidth}
    \includegraphics[width=1.0\linewidth]{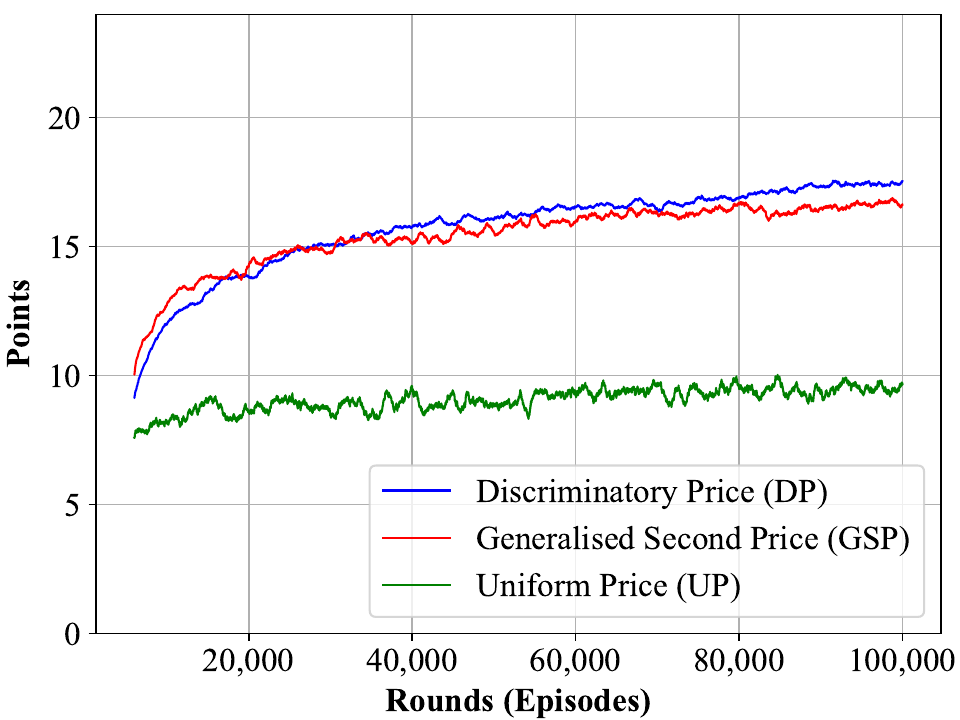}
    \caption{Auction revenue\label{fig:Revenue_7_40_1-10_4-10}}
    \end{subfigure}%
     \begin{subfigure}[t]{.48\textwidth}
    \includegraphics[width=1.0\linewidth]{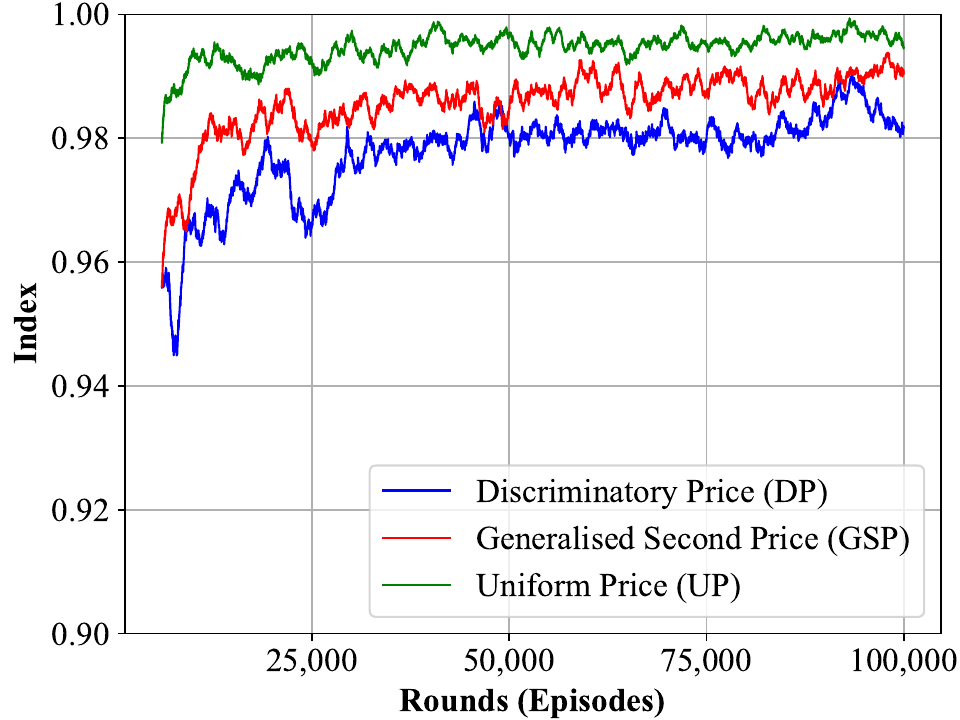}
    \caption{Auction efficiency\label{fig:Efficiency_7_40_1-10_4-10}}
    \end{subfigure}%
    \hfill\vspace{2mm}
    \begin{subfigure}[t]{.48\textwidth}
    \includegraphics[width=1.0\linewidth]{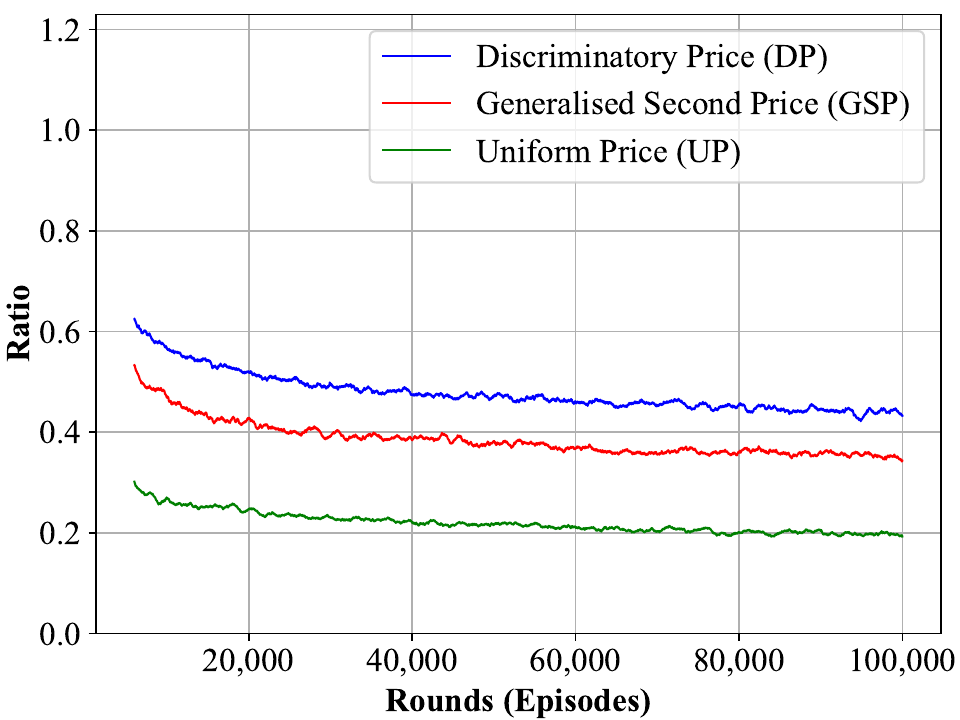}
    \caption{\centering Learning ratio: Average of 7 Q-learning agents\label{fig:Agents_7_40_1-10_4-10}}
    \end{subfigure}%
    \caption{7 Agents, Knapsack capacity of 40, item values in the range [1,10] and item sizes in the range of [4,10]\label{fig:All_7_40_1-10_4-10}}
\end{figure*}

\begin{figure*}[ht!]
    \captionsetup[sub]{font=scriptsize,labelfont={bf,sf}}
    \centering
    \begin{subfigure}[t]{.48\textwidth}
    \includegraphics[width=1.0\linewidth]{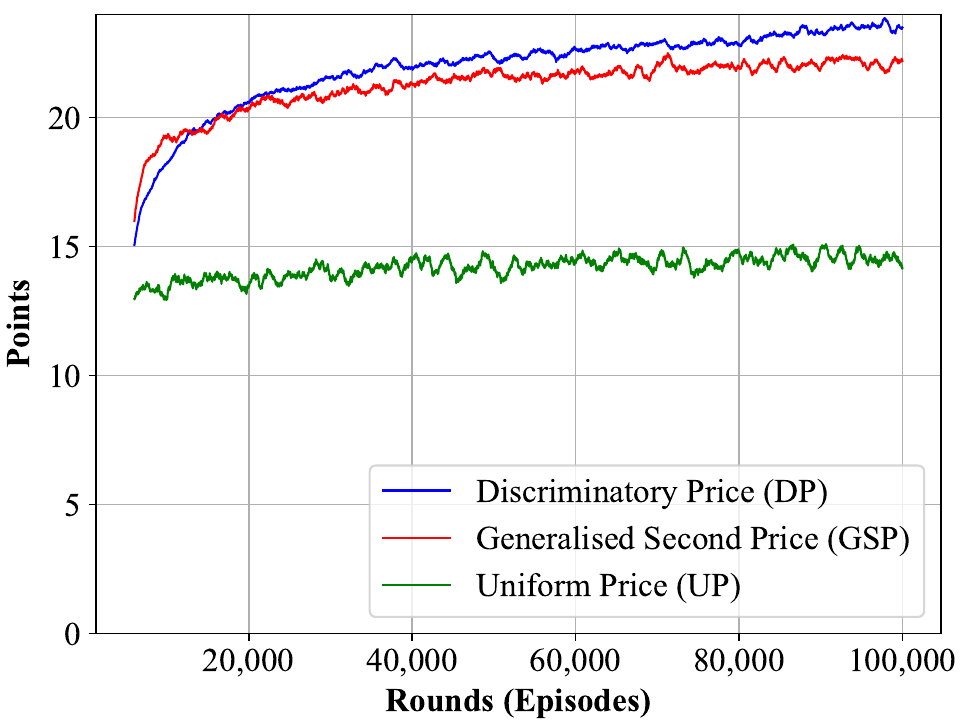}
    \caption{Auction revenue\label{fig:Revenue_10_30_1-10_1-10}}
    \end{subfigure}%
     \begin{subfigure}[t]{.48\textwidth}
    \includegraphics[width=1.0\linewidth]{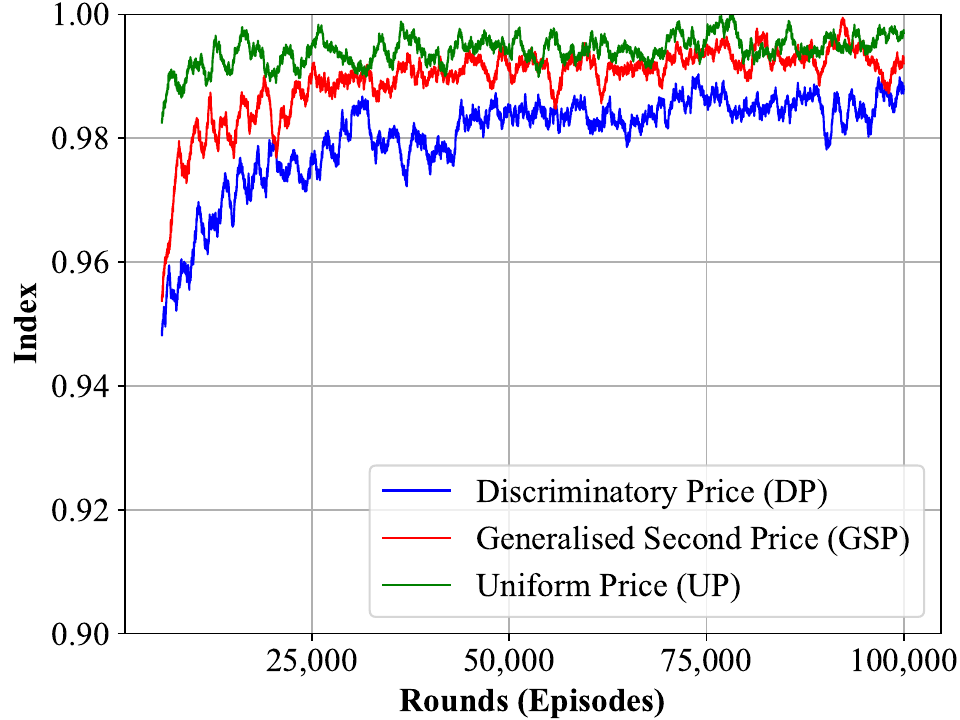}
    \caption{Auction efficiency\label{fig:Efficiency_10_30_1-10_1-10}}
    \end{subfigure}%
    \hfill\vspace{2mm}
    \begin{subfigure}[t]{.48\textwidth}
    \includegraphics[width=1.0\linewidth]{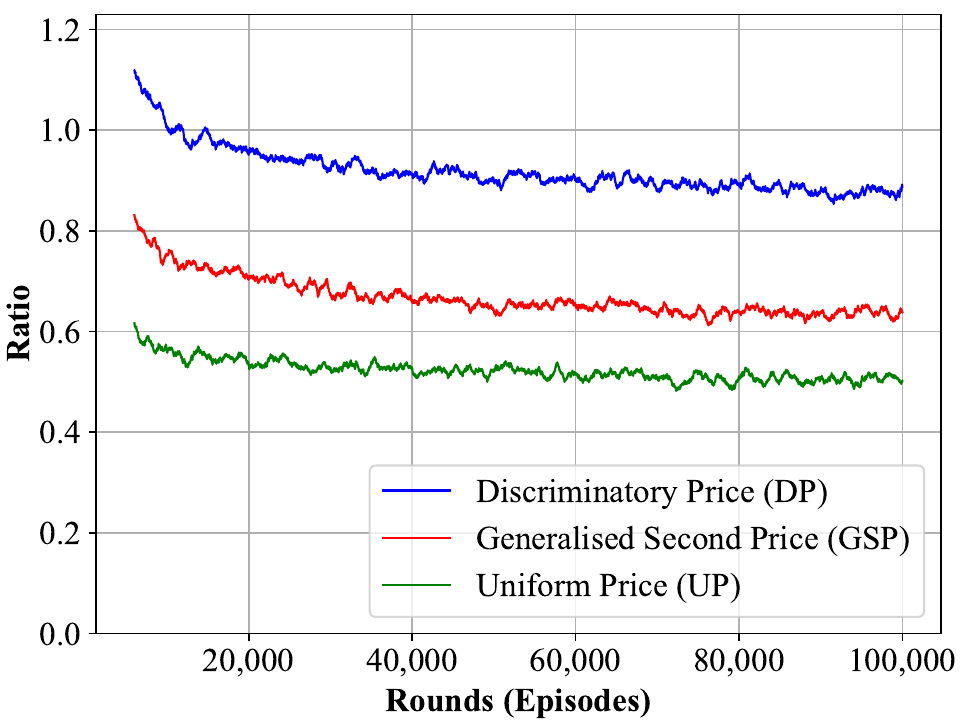}
    \caption{\centering Learning ratio: Average of 10 Q-learning agents\label{fig:Agents_10_30_1-10_1-10}}
    \end{subfigure}%
    \caption{10 Agents, Knapsack capacity of 30, item values in the range [1,10] and item sizes in the range of [1,10]\label{fig:All_10_30_1-10_1-10}}
\end{figure*}

\begin{figure*}[ht!]
    \captionsetup[sub]{font=scriptsize,labelfont={bf,sf}}
    \centering
    \begin{subfigure}[t]{.48\textwidth}
    \includegraphics[width=1.0\linewidth]{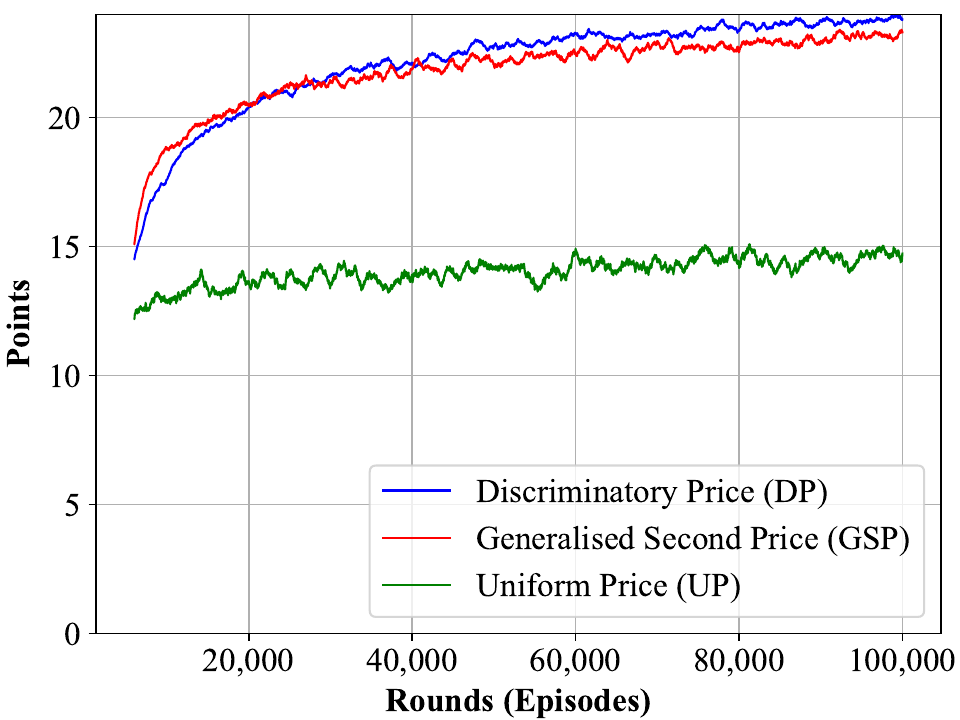}
    \caption{Auction revenue\label{fig:Revenue_10_36_1-10_1-10}}
    \end{subfigure}%
     \begin{subfigure}[t]{.48\textwidth}
    \includegraphics[width=1.0\linewidth]{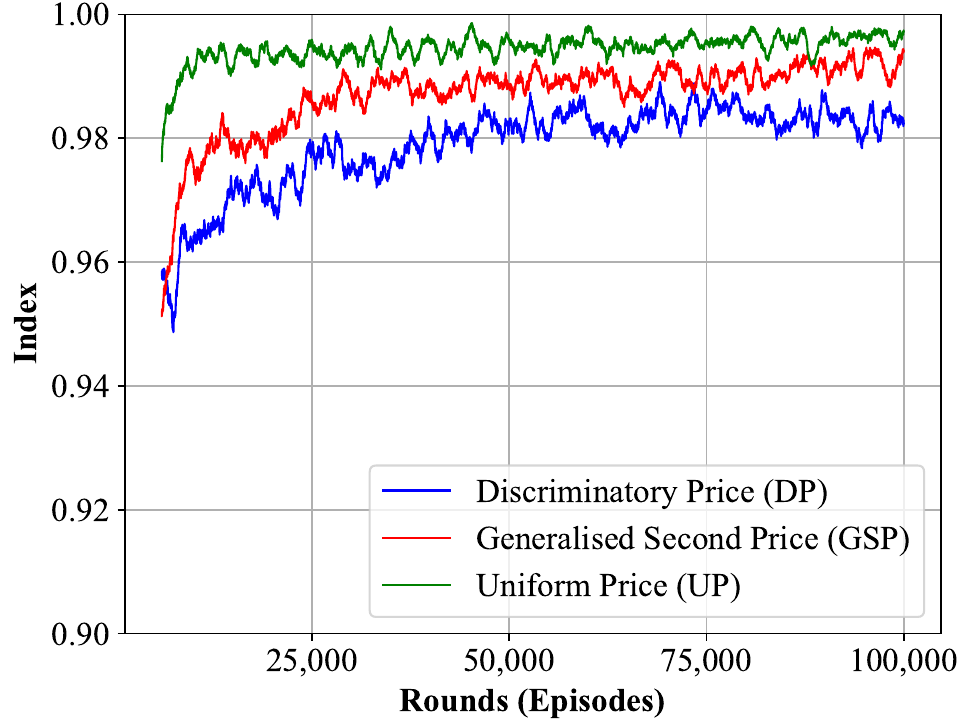}
    \caption{Auction efficiency\label{fig:Efficiency_10_36_1-10_1-10}}
    \end{subfigure}%
    \hfill\vspace{2mm}
    \begin{subfigure}[t]{.48\textwidth}
    \includegraphics[width=1.0\linewidth]{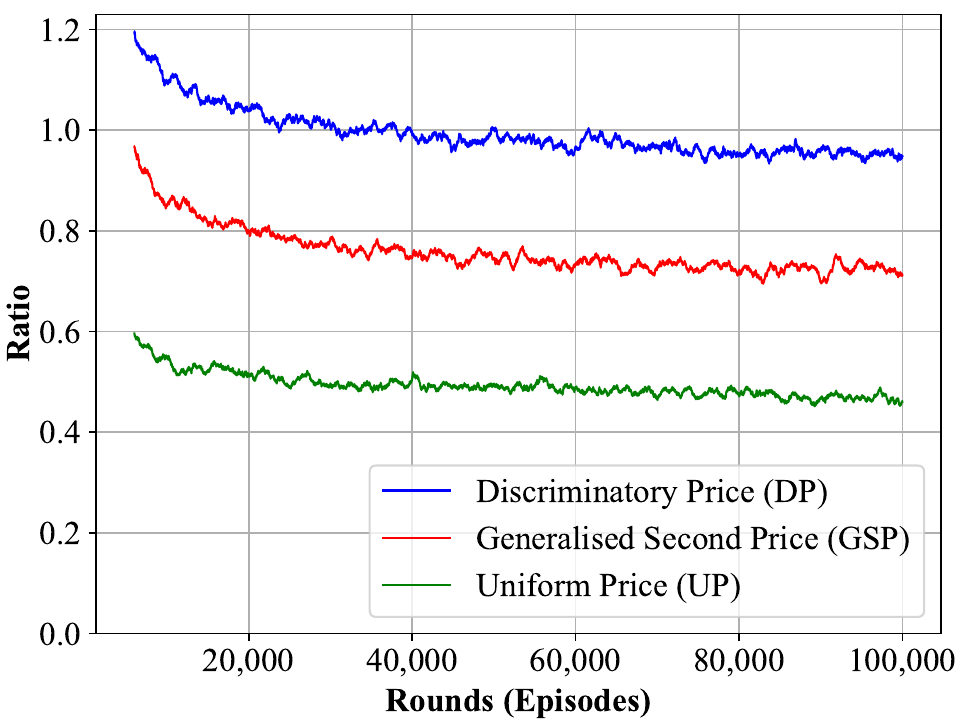}
    \caption{\centering Learning ratio: Average of 10 Q-learning agents\label{fig:Agents_10_36_1-10_1-10}}
    \end{subfigure}%
    \caption{10 Agents, Knapsack capacity of 36, item values in the range [1,10] and item sizes in the range of [1,10]\label{fig:All_10_36_1-10_1-10}}
\end{figure*}

\begin{figure*}[ht!]
    \captionsetup[sub]{font=scriptsize,labelfont={bf,sf}}
    \centering
    \begin{subfigure}[t]{.48\textwidth}
    \includegraphics[width=1.0\linewidth]{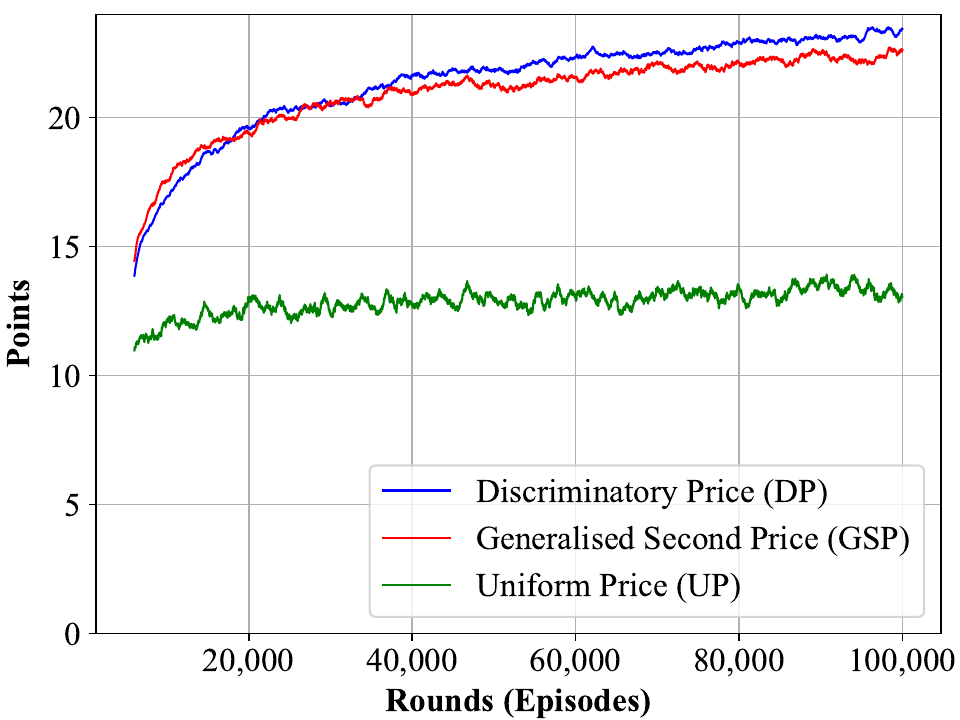}
    \caption{Auction revenue\label{fig:Revenue_10_40_1-10_1-10}}
    \end{subfigure}%
     \begin{subfigure}[t]{.48\textwidth}
    \includegraphics[width=1.0\linewidth]{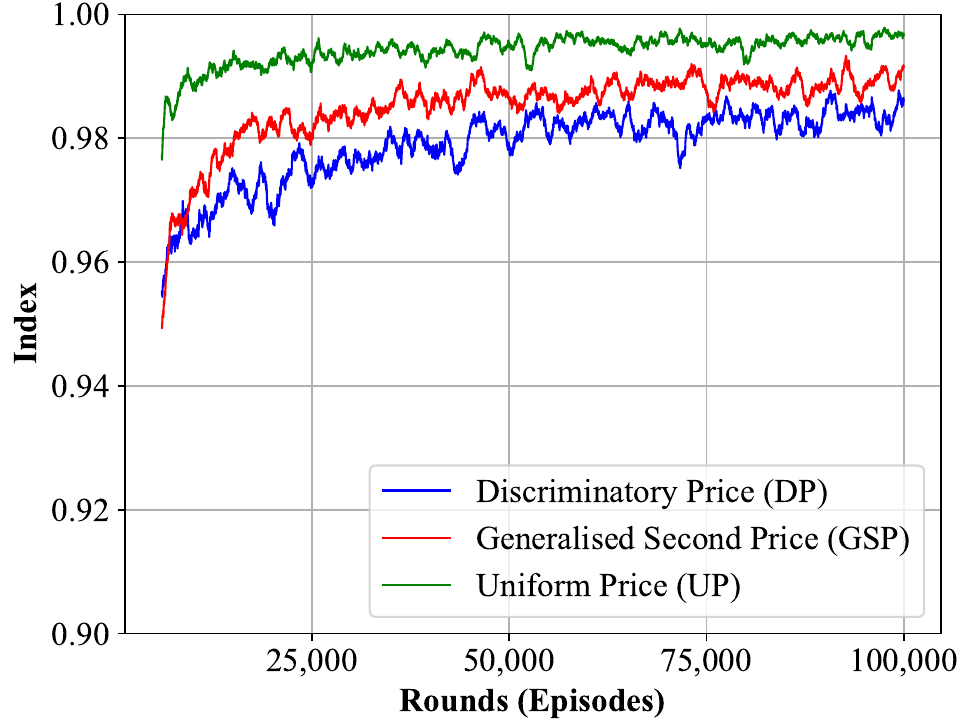}
    \caption{Auction efficiency\label{fig:Efficiency_10_40_1-10_1-10}}
    \end{subfigure}%
    \hfill\vspace{2mm}
    \begin{subfigure}[t]{.48\textwidth}
    \includegraphics[width=1.0\linewidth]{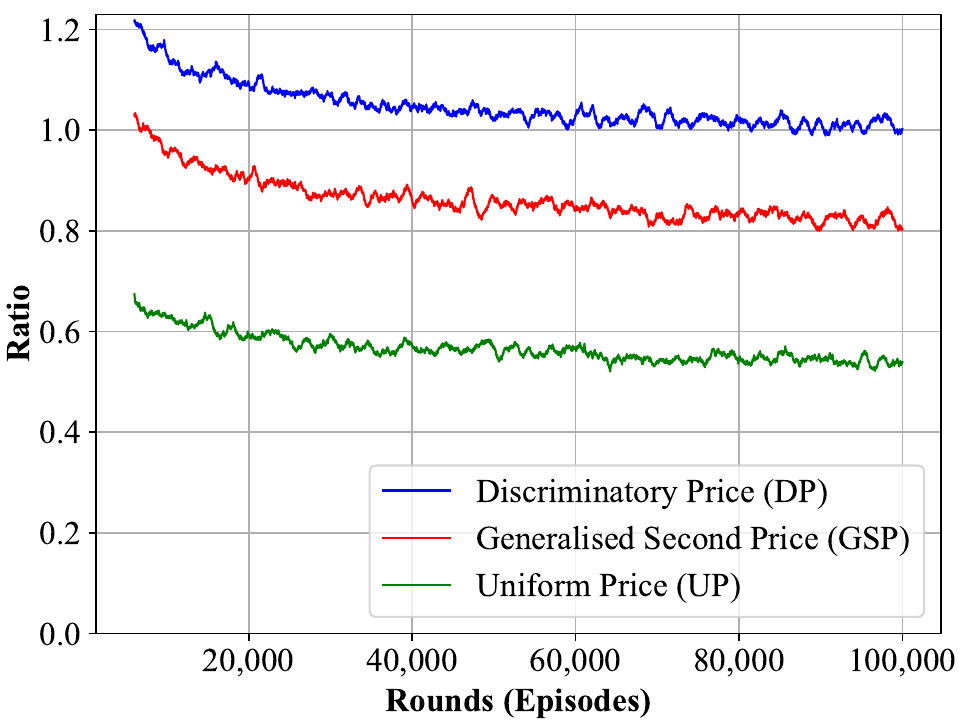}
    \caption{\centering Learning ratio: Average of 10 Q-learning agents\label{fig:Agents_10_40_1-10_1-10}}
    \end{subfigure}%
    \caption{10 Agents, Knapsack capacity of 40, item values in the range [1,10] and item sizes in the range of [1,10]\label{fig:All_10_40_1-10_1-10}}
\end{figure*}


\section{Appendix A: Details of the experiment}

\clearpage
\bibliographystyle{econ}
\bibliography{main}
\end{document}